\documentclass[aps,twocolumn,prd,showpacs,showkeys,preprintnumbers,superscriptaddress,nobibnotes,floatfix,longbibliography]{revtex4-1}

\pdfoutput=1
\usepackage{amsmath,amsfonts,amssymb,mathrsfs,graphicx,color,longtable,hyperref}

\newcommand{\ie}{{\it i.e.}}

\newcommand{\eg}{{\it e.g.}}

\newcommand{\etc}{{\it etc.}}
\newcommand{\eq}{Eq.}

\newcommand{\fig}{Fig.}
\newcommand{\Fig}{Fig.}

\newcommand{\Ref}{Ref.}
\newcommand{\Refs}{Refs.}
\newcommand{\Sec}{Section}
\newcommand{\Secs}{Sections}
\newcommand{\App}{Appendix}

\newcommand{\equ}[1]{\eq~(\ref{equ:#1})}
\newcommand{\figu}[1]{\fig~\ref{fig:#1}}

\newcommand{\bi}{\begin{itemize}}
\newcommand{\ei}{\end{itemize}}

\begin{document}

\title{Testing Decay of Astrophysical Neutrinos with Incomplete Information}

\author{Mauricio Bustamante}
\affiliation{Center for Cosmology and AstroParticle Physics (CCAPP), Ohio State University,
        Columbus, OH 43210, USA}
\affiliation{Department of Physics, Ohio State University, Columbus, OH 43210, USA}

\author{John F.~Beacom}
\affiliation{Center for Cosmology and AstroParticle Physics (CCAPP), Ohio State University,
        Columbus, OH 43210, USA}
\affiliation{Department of Physics, Ohio State University, Columbus, OH 43210, USA}
\affiliation{Department of Astronomy, Ohio State University, Columbus, OH 43210, USA}

\author{Kohta Murase}
\affiliation{Center for Particle and Gravitational Astrophysics, Pennsylvania State University, University Park, Pennsylvania, 16802, USA}
\affiliation{Department of Physics, Pennsylvania State University, University Park, Pennsylvania, 16802, USA}
\affiliation{Department of Astronomy \& Astrophysics, Pennsylvania State University, University Park, Pennsylvania, 16802, USA
\\
{\tt \href{mailto:bustamanteramirez.1@osu.edu}{bustamanteramirez.1@osu.edu}, \href{mailto:beacom.7@osu.edu}{beacom.7@osu.edu}, \href{mailto:murase@psu.edu}{murase@psu.edu}} \\
{\tt \footnotesize \href{http://orcid.org/0000-0001-6923-0865}{0000-0001-6923-0865}, \href{http://orcid.org/0000-0002-0005-2631}{0000-0002-0005-2631}, \href{http://orcid.org/0000-0002-5358-5642}{0000-0002-5358-5642} \smallskip}}

\date{January 25, 2017}

\begin{abstract}

Neutrinos mix and have mass differences, so decays from one to another must occur.  But how fast?  The best direct limits on non-radiative decays, based on solar and atmospheric neutrinos, are weak, $\tau \gtrsim 10^{-3}$ s ($m$/eV) or much worse.  Greatly improved sensitivity, $\tau \sim 10^3$ s ($m$/eV), will eventually be obtained using neutrinos from distant astrophysical sources, but large uncertainties --- in neutrino properties, source properties, and detection aspects --- do not allow this yet.  However, there is a way forward now.  We show that IceCube diffuse neutrino measurements, supplemented by improvements expected in the near term, can increase sensitivity to $\tau \sim 10$ s ($m$/eV) for all neutrino mass eigenstates.  We provide a roadmap for the necessary analyses and show how to manage the many uncertainties.  If limits are set, this would definitively rule out the long-considered possibility that neutrino decay affects solar, atmospheric, or terrestrial neutrino experiments.

\end{abstract}


\maketitle

\section{Introduction}\label{section:Introduction}

No symmetry protects heavier neutrino mass eigenstates from decaying into lighter ones, though the stability of the lightest neutrino is presumably guaranteed by lepton number conservation.  However, the expected decay lifetimes in the Standard Model, minimally extended to include neutrino masses, are $\gtrsim 10^{43}$~s~\cite{Pal:1981rm, Hosotani:1981mq, Nieves:1982bq}, which are so long as to be irrelevant.  

Detection of neutrino decay would therefore signal new physics.  Decay rates can be dramatically enhanced by couplings to new particles, especially those with masses small enough to be among the decay products.  Searches for new physics at low masses with neutrinos are complementary to searches for new particles at high masses with colliders.  We focus on the challenging case of non-radiative decays, {\it i.e.}, with no final-state photons.

Neutrino decay rates depend on the factor
\begin{equation}
\exp\left(-\frac{t}{\gamma \tau}\right) = \exp\left(-\frac{L}{E} \times \frac{m}{\tau}\right) \,,
\end{equation}
where $t$ is the elapsed time since production, $L \approx t$ is the traveled distance, $\tau$ is the lifetime, and $\gamma \equiv E/m$ is the Lorentz boost, with $E$ and $m$ the energy and mass; we have taken $c = 1$.  This factor governs the disappearance rate of parent neutrinos and the appearance rate of possibly active daughter neutrinos.  Decay occurs between mass eigenstates with well-defined lifetimes.  However, neutrinos are usually produced and detected in flavor eigenstates, so care is needed to probe decay in the presence of large mixing.  Sensitivity to neutrino decay depends on the precision of a flux prefactor, not shown above, and the deviation of the exponential from unity.

To test long lifetimes, one is driven to the large distances of astrophysical sources, like those of the IceCube neutrinos\ \cite{Aartsen:2013bka,Aartsen:2013jdh,Aartsen:2013eka,Aartsen:2014gkd,Aartsen:2015knd,Aartsen:2015rwa,Aartsen:2015xup,Aartsen:2016xlq}. 
Decay has even been invoked to explain features of the IceCube signal\ \cite{Baerwald:2012kc,Pakvasa:2012db,Dorame:2013lka}.
However, for astrophysical neutrinos, decay seemingly must be tested in an absolute sense, using theoretical knowledge of the source flux, unlike for, say, atmospheric neutrinos, where decay can be tested in a relative sense, comparing upgoing and downgoing rates.  This problem can be solved by using the flavor composition --- the ratios of $\nu_e + \bar{\nu}_e$, $\nu_\mu + \bar{\nu}_\mu$, and $\nu_\tau + \bar{\nu}_\tau$ to the total flux --- to formulate a relative test~\cite{Pakvasa:1981ci,Beacom:2002vi,Barenboim:2003jm,Beacom:2003nh,Beacom:2003zg,Meloni:2006gv,Maltoni:2008jr,
Bustamante:2010nq,Baerwald:2012kc,Pagliaroli:2015rca,Bustamante:2015waa,Huang:2015flc,Shoemaker:2015qul}.

Even so, discussions of testing neutrino decay typically make strong assumptions, including the following:
\begin{itemize}

\vspace{-0.2cm}

\item {\bf Neutrino properties:} \\
Daughter neutrino properties are known. \\
Decay modes are known. \\
Mixing parameters are known.

\vspace{-0.2cm}

\item {\bf Source properties:} \\
Distances to the source(s) are known. \\
Energy spectra at the source(s) are known. \\
Flavor ratios at the sources(s) are known.

\vspace{-0.2cm}

\item {\bf Detection aspects:} \\
Energy is measured well for each neutrino. \\
Flavor is measured well for each neutrino. \\
Negligible contribution from background events. 

\vspace{-0.2cm}

\end{itemize}
{\it At present, none of these conditions are fully met.}  Despite this, we show that interesting sensitivity, robust against uncertainties, can be obtained with IceCube in the near term.  We focus on methods and order-of-magnitude estimates, leaving details to experimental studies.

\begin{figure*}[t]
 \begin{center}
  \includegraphics[clip=false, trim = 0.9cm 1.9cm 0 0, width=\columnwidth]{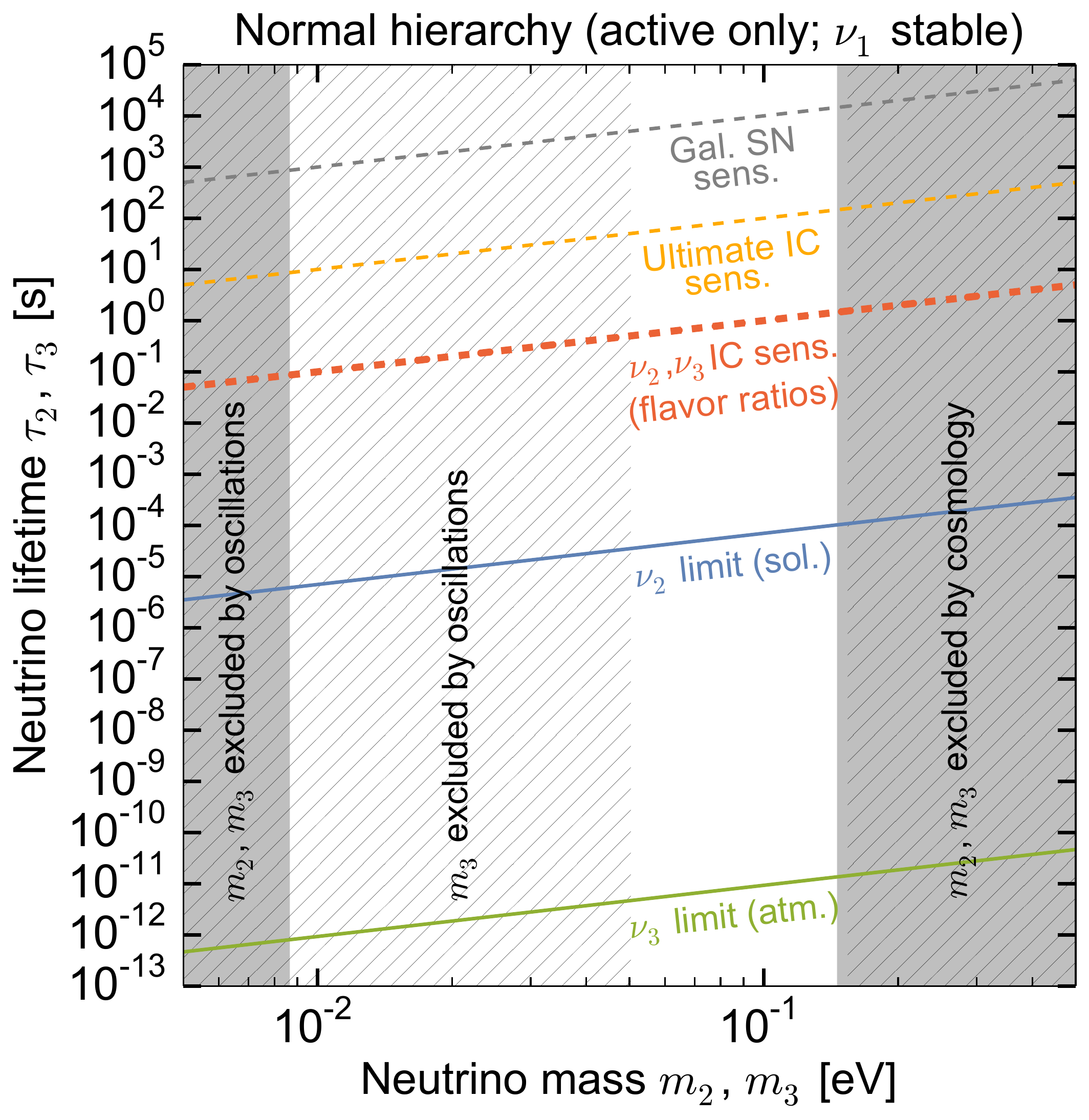}\;\;\;\;\;
  \includegraphics[clip=false, trim = 0.9cm 1.9cm 0 0, width=\columnwidth]{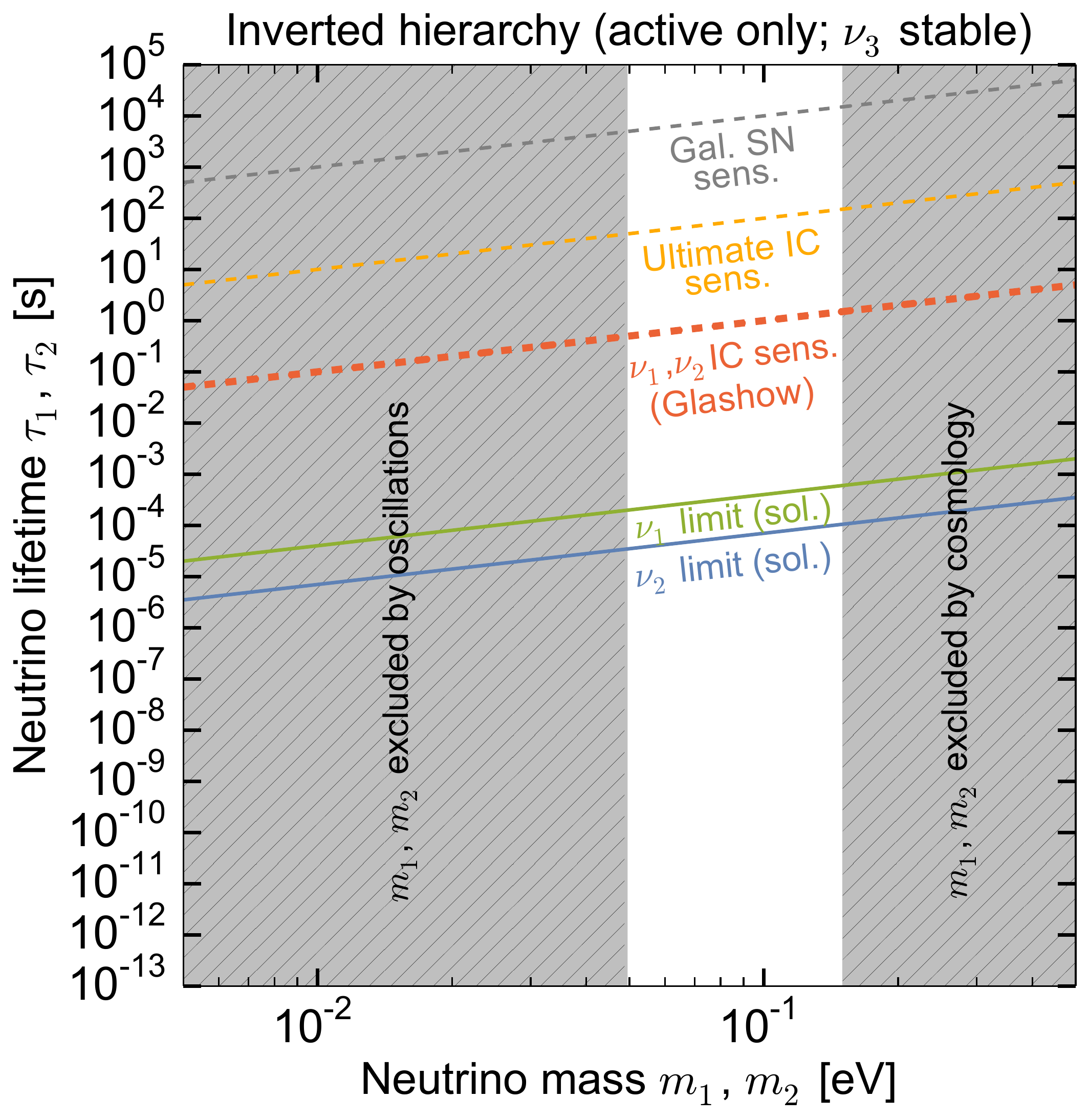}
 \end{center}
 \caption{\label{fig:mass-vs-lifetime}Constraints on neutrino masses and lifetimes, as labeled and discussed in the text, with hatched gray disallowed, hatched white allowed only for some eigenstates, and non-hatched white allowed for all.  Solid lines are lower limits.  The thick red dashed lines indicate the sensitivity estimates of this paper.  {\it Left:} Normal hierarchy.  {\it Right:} Inverted hierarchy.}
\end{figure*}

This paper is organized as follows.  In \Sec\ \ref{section:Lifetimes}, we review neutrino lifetime limits and sensitivities.  In \Secs\ \ref{section:ManagingNeutrinoProperties}, \ref{section:ManagingSourceProperties}, and \ref{section:ManagingDetection}, we show that uncertainties in neutrino properties, uncertainties in source properties, and detection aspects are manageable.  In \Sec\ \ref{section:EstimatingLimits}, we estimate lifetime sensitivities achievable by IceCube. In \Sec\ \ref{section:SummaryConclusions}, we summarize and conclude.


\section{Overview of neutrino lifetime limits and sensitivities}\label{section:Lifetimes}

Figure \ref{fig:mass-vs-lifetime} shows present limits and future sensitivities on lifetimes and masses of mass eigenstates $\nu_i$ ($i = 1,2,3$).  
(Here and below, $\nu_i$ stands for $\nu_i + \bar{\nu}_i$ and $\nu_\alpha$ stands for $\nu_\alpha + \bar{\nu}_\alpha$ ($\alpha = e, \mu, \tau$), unless otherwise indicated.)
Since the neutrino mass hierarchy is unknown, we consider the two possibilities.  In the normal hierarchy (NH), $\nu_2$ and $\nu_3$ are unstable and heavier than $\nu_1$, which is stable.  In the inverted hierarchy (IH), $\nu_1$ and $\nu_2$ are unstable and heavier than $\nu_3$, which is stable.  (We assume only three active neutrinos --- $\nu_e$, $\nu_\mu$, $\nu_\tau$, or $\nu_1$, $\nu_2$, $\nu_3$ --- and no mixing with sterile neutrinos\ \cite{Kopp:2013vaa,deGouvea:2015euy,Arguelles:2016uwb,TheIceCube:2016oqi}.)  

The allowed mass range is strikingly narrow.  Lower limits come from the squared-mass differences $\Delta m_{ij}^2 \equiv m_i^2 - m_j^2$ measured in neutrino oscillation experiments~\cite{Gonzalez-Garcia:2014bfa}.  Upper limits come from cosmological constraints on the sum of masses~\cite{Agashe:2014kda}.  We have conservatively assumed $\sum_i m_i \lesssim 0.3~\text{eV}$.  Recent work~\cite{Palanque-Delabrouille:2015pga} claims $\sum_i m_i \lesssim 0.12$ eV --- and the bounds are expected to continue improving --- which would result in even narrower allowed mass ranges.  In these plots, we considered $m_1 = 0$ for NH and $m_3 = 0$ for IH, to show the widest ranges.  Below, we discuss the implications of this narrow allowed mass range.

For detected neutrinos with known $L$ and $E$, there is nominal sensitivity to lifetimes of
\begin{equation}\label{equ:LifetimeSensitivity}
 \frac { \tau } { m }
 \sim
 10^3
 \left( \frac{ L } { \text{Gpc} } \right) 
 \left( \frac{ 100\ \text{TeV} } { E } \right) 
 \, \text{s}\ \text{eV}^{-1} \;.
\end{equation}
Though our interest is in $\tau$, only the combination $\tau/m$ is observable. How $\tau$ ($\equiv 1/\Gamma$, where $\Gamma$ is the decay rate) itself depends on the masses of parent and daughter neutrinos is model-dependent (see, \eg, \Ref\ \cite{Beacom:2002cb}).  Here we focus on lifetime sensitivity that is direct (based on neutrino detection), as this gives the greatest generality.  Below, we remark on indirect cosmological limits that apply to certain scenarios. 

In the NH, the limit for $\nu_2$ comes from solar neutrino experiments, $\tau_2/m_2 \gtrsim 7 \cdot 10^{-4}$ s eV$^{-1}$ \cite{Berryman:2014qha}; for $\nu_3$, since it is not probed by solar neutrinos, the limit comes from atmospheric and long-baseline experiments, $\tau_3/m_3 \gtrsim 9 \cdot 10^{-11}$ s eV$^{-1}$ \ \cite{GonzalezGarcia:2008ru} (see also \Ref\ \cite{Gomes:2014yua}).  In the IH, limits for $\nu_1$ and $\nu_2$ come from solar neutrino experiments\ \cite{Berryman:2014qha}: $\tau_1/m_1 \gtrsim 4 \cdot 10^{-3}$ s eV$^{-1}$ and $\tau_2/m_2 \gtrsim 7 \cdot 10^{-4}$ s eV$^{-1}$ (see also \Refs\ \cite{Joshipura:2002fb,Beacom:2002cb,Bandyopadhyay:2002xj,Picoreti:2015ika}).  Though weak, these are the best limits we have.

The detection of neutrinos with tens of MeV from supernova 1987A\ \cite{Hirata:1987hu,Bionta:1987qt,1987ESOC...26..237A}, located $\sim$ 50 kpc away, could naively be used to set a lifetime limit.  However, due to large uncertainties in emission and neutrino mixing, and to the detection of only one flavor, no robust limit has been demonstrated in a three-neutrino scenario.
Instead, we show the estimated ``Galactic supernova sensitivity'' of $\tau/m \gtrsim 10^5$ s eV$^{-1}$ that could be reached by detecting neutrinos of 10 MeV from a supernova 10 kpc away.  
Detection of more than one flavor of supernova neutrinos in next-generation experiments like Hyper-Kamiokande\ \cite{Abe:2011ts} and DUNE\ \cite{Ankowski:2016lab} could help determine the explosion mechanism and, with that, improve this figure.
 
IceCube recently discovered a diffuse flux of astrophysical neutrinos between 25 TeV and 10 PeV\ \cite{Aartsen:2013bka,Aartsen:2013jdh,Aartsen:2013eka,Aartsen:2014gkd,Aartsen:2015knd,Aartsen:2015rwa,Aartsen:2015xup,Aartsen:2016xlq} (for reviews, see, {\it e.g.}, \Refs\ \cite{Anchordoqui:2013dnh,Murase:2014tsa,Ishihara:2015xuv}).  No point sources have been identified\ \cite{Aartsen:2016oji}.  The ``ultimate IceCube sensitivity'' of $\tau/m \gtrsim 10^3$ s eV$^{-1}$ could be reached by detecting neutrinos of 100 TeV from sources 1 Gpc away, if all conditions from \Sec\ \ref{section:Introduction} are met.  But, with current and near-future data, this is unfeasible. 

We show that the situation is really not dire: already now, IceCube should be able to achieve sensitivities that approach its ultimate sensitivity.  Our new estimated sensitivity ``IC sens.''\ of $\tau/m \gtrsim 10$ s eV$^{-1}$ far outperforms existing limits.  Below, we explain how this is derived.


\section{Managing uncertainties in neutrino properties}\label{section:ManagingNeutrinoProperties}


\subsection{More general treatment of decay modes}\label{section:NewPerspective}

Until recently, studies of decay focused on the triply-degenerate mass scenario ($m_1 \approx m_2 \approx m_3$), motivated by the former large allowed mass range.  This forced parent and daughter neutrinos to have  almost equal masses, the daughter to carry almost the full energy of the parent, and any additional decay product to be massless or very light.  
As a result, the effects depend on whether or not neutrino daughters are active (see \Ref\ \cite{Beacom:2002cb} for a comparison using solar neutrinos), and on the particular decay mechanism\ \cite{Chikashige:1980qk,Gelmini:1982rr,Tomas:2001dh,Hannestad:2005ex,Zhou:2007zq,Chen:2007zy,Li:2007kj}.
Past work focused on decay to a daughter neutrino and a massless particle.  Such scenarios are strongly restricted by indirect limits from cosmology\ \cite{Hannestad:2005ex,Serpico:2007pt} (or astrophysics\ \cite{Kachelriess:2000qc,Farzan:2002wx}). 
 
Figure \ref{fig:mass-vs-lifetime} reveals a new perspective, spurred by recent progress in measuring neutrino masses.  For a fixed $\Delta m_{ij}^2$, the mass difference $m_i - m_j = \Delta m_{ij}^2 / ( m_i + m_j )$ rises at lower masses.  Because the allowed mass range is low and narrow, that implies that the triply-degenerate scenario is becoming less likely.

This makes our analysis more model-independent.  A hierarchical mass scheme, where $m_1 \ll m_2, m_3$ in the NH and $m_3 \ll m_1, m_2$ in the IH (see \Fig~1 in \Ref~\cite{Beacom:2002cb} and \Fig~8 in \Ref~\cite{Bilenky:2002aw}), opens up previously unmotivated possibilities for decay.
It allows us to consider that additional decay products can be massive (though light), and that the cosmological limits may have to be reconsidered.
Instead of the final state being a non-relativistic neutrino and an additional extreme-relativistic particle (in the center-of-mass frame of the parent neutrino), all possibilities on the kinematics are now allowed.

Further, the different kinematics also affects whether or not it is important that neutrino daughters are active.  Because of the large mass splitting between parent and daughter neutrinos, the daughter carries only a fraction of the energy of the parent, unlike for the degenerate-mass case.  Thus, for falling spectra, daughter neutrinos may be unimportant even if active.

We focus on complete decay, where all unstable neutrinos have decayed upon reaching the detector, imprinting the largest effects.  In this scenario, we do not need to consider branching ratios into individual decay modes.  For concreteness, we consider that the parent neutrino decays into the lightest neutrino and one or more extra particles.  In the NH, the neutrino component of the decays is $\nu_2, \nu_3 \to \nu_1$; in the IH, it is $\nu_1, \nu_2 \to \nu_3$.  (For a comprehensive list of alternative scenarios, see \Ref\ \cite{Maltoni:2008jr}.)

Current data is compatible with a power-law neutrino spectrum, expected from theoretical considerations.  The spectral index with complete decay would be the same as with no decay.  Since the flux normalization is {\it a priori} unknown, this allows us to simplify the discussion.  If daughter neutrinos are sterile (``invisible''), they will not contribute at all.  If daughter neutrinos are active (``visible''), they will contribute some fraction of the parent energy, inheriting the same spectral index.  Therefore, with limited data, we are insensitive to whether or not decay products contribute to the detected flux. 

We forgo looking for the transition between no decay and complete decay, which would help as an additional observable.  More sophisticated analyses, with more data, could do that, by combining flavor and spectral information\ \cite{Shoemaker:2015qul}.  If there is a feature in the neutrino spectrum due to decay --- or from a cut-off in the emission spectrum --- then the properties of daughter neutrinos should be considered more carefully.


\subsection{Managing uncertainties in neutrino mixing}

\begin{figure}[t!]
 \centering
 \includegraphics[width=\columnwidth,clip=true,trim=0 0.3cm 0 0.8cm]{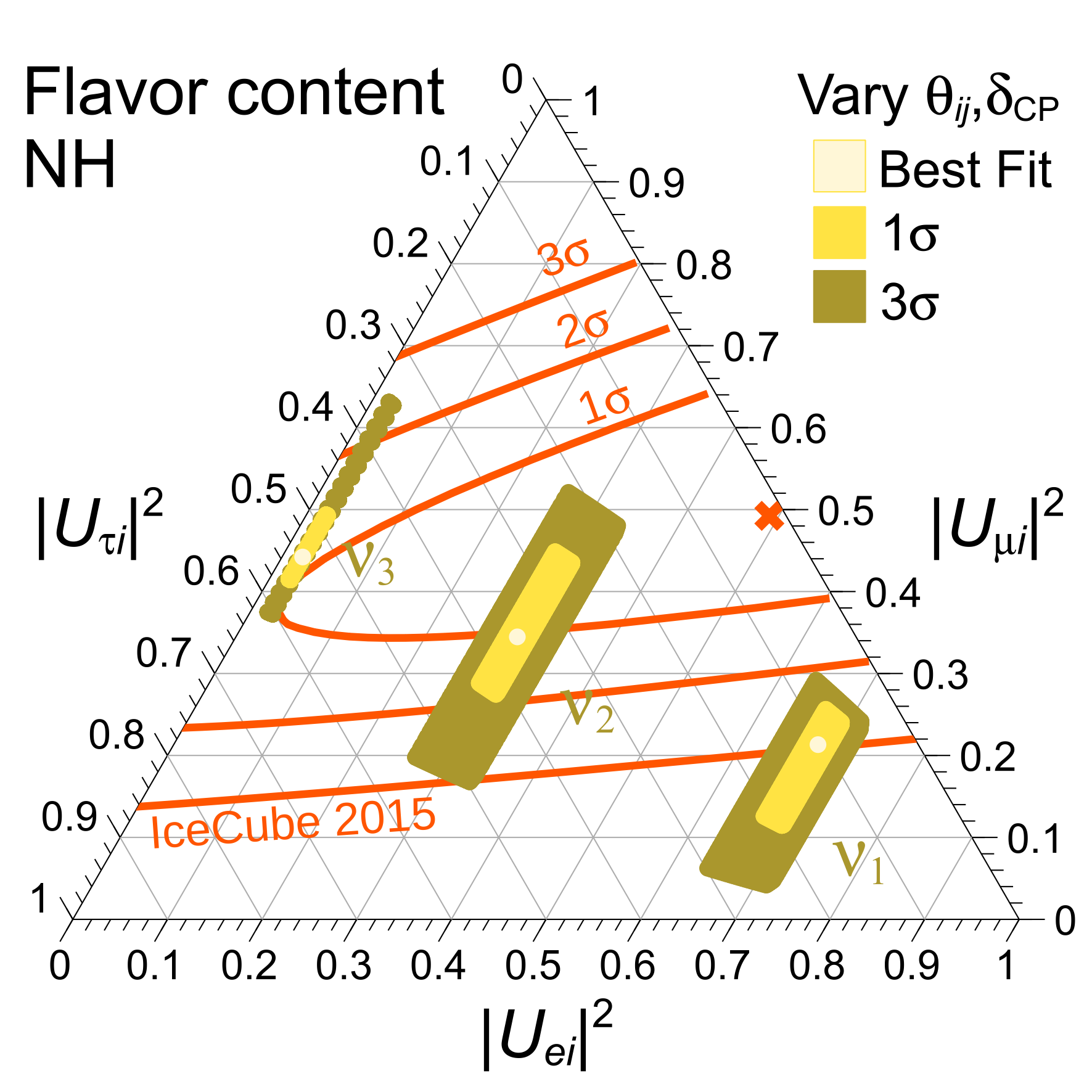}
 \caption{\label{fig:ternary-plot-flavor-content-NH}Flavor content of mass eigenstates $\nu_1$, $\nu_2$, and $\nu_3$, in the NH (results for the IH are very similar\ \cite{Bustamante:2015waa}). The regions are generated using the best-fit value of the mixing parameters (light yellow), and their $1\sigma$ (darker) and $3\sigma$ (darkest) uncertainty ranges from \Ref~\cite{Gonzalez-Garcia:2014bfa}. IceCube astrophysical flavor composition measurements~\cite{Aartsen:2015knd} are shown.  Values are read parallel to their ticks.  Figure modified from \Fig\ 1 in \Ref\ \cite{Bustamante:2015waa}.}
\end{figure}

Decay occurs between mass eigenstates, but neutrino detectors are sensitive to flavor states for the dominant detection channel of neutrino-nucleon charged-current interactions.  Mixing between the two bases is large.  It is commonly represented by the PMNS matrix $U$ and parametrized by three angles --- $\theta_{12} \approx 34^\circ$, $\theta_{23} \approx 45^\circ$, and $\theta_{13} \approx 9^\circ$ --- and one CP-violation phase $\delta_\text{CP}$, still unconstrained.  Uncertainties in the angles are small and shrinking, but not negligible\ \cite{Gonzalez-Garcia:2014bfa} (also, \Refs\ \cite{Forero:2014bxa,Gonzalez-Garcia:2015qrr}).  Next, we show that present uncertainties are not an obstacle to testing decay.

Figure\ \ref{fig:ternary-plot-flavor-content-NH} shows the regions of flavor content $\lvert U_{\alpha i} \rvert^2$ of the mass eigenstates $\nu_i$, generated by varying the mixing parameters within their allowed ranges, from \Ref\ \cite{Bustamante:2015waa}.  They are clearly separable, which means that a flux of pure $\nu_1$ and a flux of pure $\nu_3$ would be distinguishable, barring detection aspects.  Results for NH and IH are similar; see Fig.\ A.1 in \Ref\ \cite{Bustamante:2015waa}.

The size of the short sides of the regions in \figu{ternary-plot-flavor-content-NH} is determined by the small uncertainties in $\theta_{12}$ and $\theta_{13}$; the size of the long sides is determined by the larger uncertainties in $\theta_{23}$ and $\delta_\text{CP}$.  Future reduced uncertainties in the mixing angles will shrink the flavor-content regions in \figu{ternary-plot-flavor-content-NH}, sharpening the separation between them; see Fig.\ C.1 in \Ref\ \cite{Bustamante:2015waa} and Figs.\ 5 and 8 in \Ref\ \cite{Shoemaker:2015qul}.


\subsection{Summary}

Because the neutrino flux --- a power law, as indicated by data --- has a normalization that is {\it a priori} unknown, under complete decay there is no sensitivity to whether or not daughter neutrinos are active and to what fraction of the parent neutrino energy they receive.  
As a result, there is no sensitivity to different decay modes.
This lack of sensitivity is exploited here to estimate model-independent sensitivities to neutrino lifetime.
Additionally, uncertainties in mixing parameters are small enough for the flavor-content regions of $\nu_1$ and $\nu_3$, corresponding to complete decay in the NH and IH, to be well separated.


\section{Managing uncertainties in source properties}\label{section:ManagingSourceProperties}


\subsection{Introducing cosmological effects on decay}\label{section:NeutrinoDecayReshift}

So far in our discussion, we have neglected two physical effects: the scaling of energy with redshift due to cosmological expansion and the dependence on redshift of the time traveled by the neutrino, as measured by its own clock, via the look-back distance\ \cite{Baerwald:2012kc}.  Taking them into account, the fraction of $\nu_i$, emitted by a source with redshift $z$, that remains upon reaching Earth, is
\begin{equation}\label{equ:DiDefinition}
 D\left(E_0,z,\tau/m\right) = \left[ \mathcal{Z}\left(z\right) \right]^{- \frac{m}{\tau} \cdot \frac{L_H}{E_0}} \;,
\end{equation}
where $E_0$ is the received neutrino energy, while the energy at emission was $E_0 \left( 1 + z \right)$, and $L_H \approx 3.89$ Gpc is the Hubble length.  The redshift-dependent part is $\mathcal{Z}\left(z\right) \simeq a + b e^{-cz}$, with $a \approx1.71$, $b = 1-a$, and $c \approx 1.27$ for a $\Lambda$CDM cosmology with $\Omega_m = 0.27$ and $\Omega_\Lambda = 0.73$.  For stable eigenstates, $D = 1$; for unstable ones, $D < 1$. If $D \ll 1$ for all unstable neutrinos, decay is complete.  \equ{DiDefinition} was first derived in \Ref~\cite{Baerwald:2012kc} (see \Ref\ \cite{Wagner:1997vn} for a related application to neutrino oscillations).

\begin{figure}[t]
 \begin{center}
  \includegraphics[width=\columnwidth]{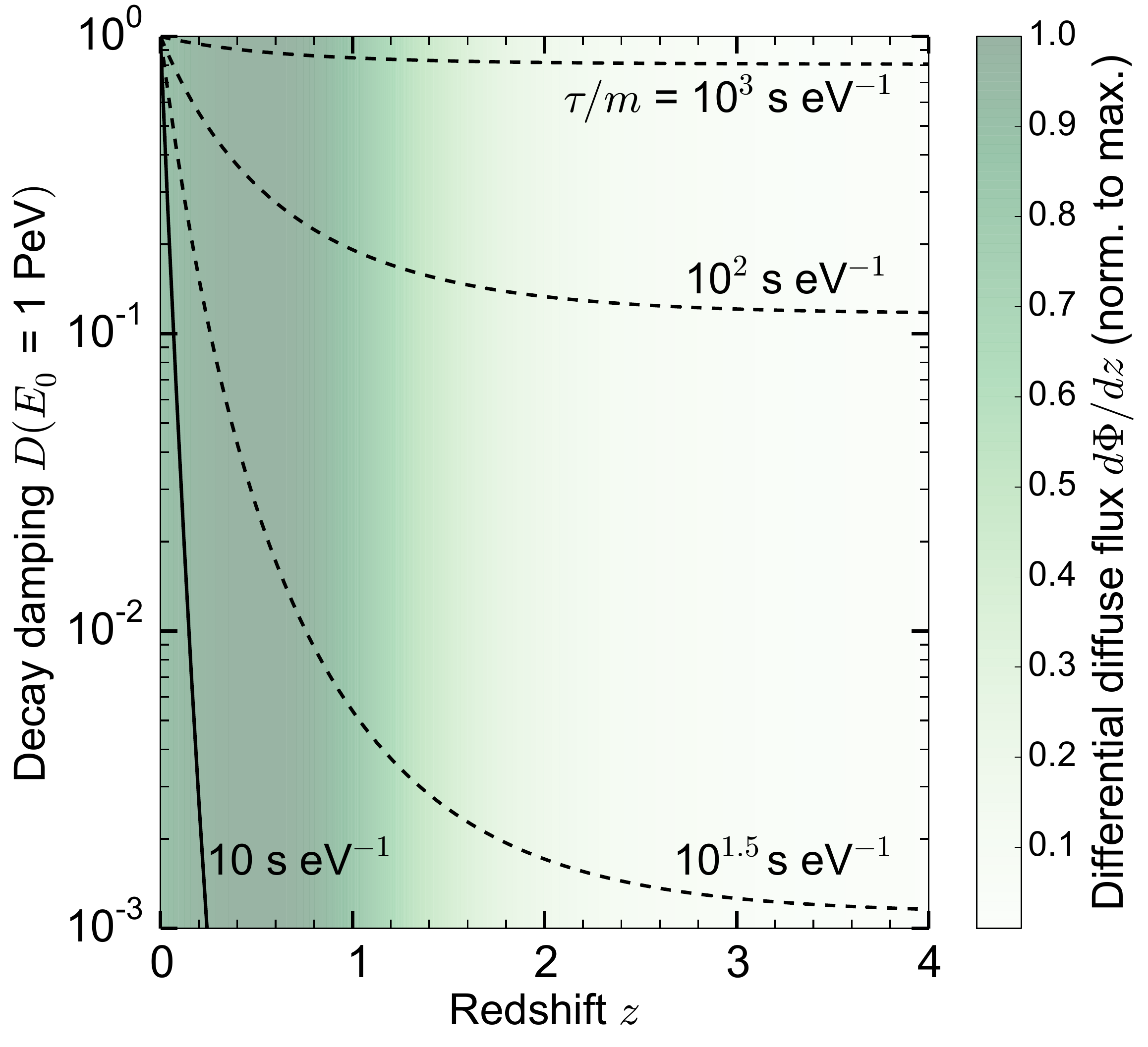}
 \end{center}
 \caption{\label{fig:D-vs-redshift}Decay damping $D$ as a function of redshift, for a fixed received neutrino energy $E_0 = 1$ PeV and different values of lifetime $\tau/m$.  The background shading is darker the higher the differential diffuse flux, assuming $\gamma = 2.50$ (see \figu{differential-flux-vs-redshift}).}
\end{figure}

Figure \ref{fig:D-vs-redshift} shows the cumulative effect of decay, for a fixed received energy of 1 PeV. 
For a lifetime of $10^3$ s eV$^{-1}$, $D \approx 1$ for the most important redshifts, which means that reaching the ultimate IceCube sensitivity will be challenging.  
For our projected sensitivity of 10 s eV$^{-1}$, decay would leave a strong imprint, since it would be complete ($D \ll 1$) for all but local sources. 

Figure \ref{fig:D-vs-lifetime} shows how the decay damping varies with lifetime, for different values of received energy in the IceCube range.  For a lifetime of 10 s eV$^{-1}$, decay is essentially complete for most of the range.


\subsection{Introducing decay in the flavor composition}\label{section:FlavorCompositionDecay}

Decay occurs along flavor oscillations. However, they have very different length scales.  Neutrinos either leave a source as incoherent mass eigenstates due to matter effects or nearly immediately become so with vacuum mixing due to the short oscillation length, $\sim 10^{-15}$~Mpc~$(E/\text{TeV})$.  After a few oscillation lengths, the $\nu_\alpha \to \nu_\beta$ flavor-transition probability averages out to $P_{\alpha\beta} = \sum_i \left\vert U_{\alpha i} \right\vert^2 \left\vert U_{\beta i} \right\vert^2$.  The decay length is orders of magnitude larger, $\sim 0.01~\text{Mpc}~( \tau / \text{s} ) / ( m / \text{eV})  ( E/\text{TeV} ) $.

With decay, the flavor-transition probability becomes energy- and redshift-dependent: $P_{\alpha\beta}\left(E_0,z,\tau_i/m_i\right) = \sum_i \left\vert U_{\alpha i} \right\vert^2 \left\vert U_{\beta i} \right\vert^2 D\left(E_0,z,\tau_i/m_i\right)$.  See \App\ \ref{appendix:FlavorTransition}.

The flavor ratios of astrophysical neutrinos can reveal information about conditions at production, propagation, and detection~\cite{Barenboim:2003jm,Kashti:2005qa,Xing:2006uk,Lipari:2007su,Pakvasa:2007dc,Esmaili:2009dz,Lai:2009ke,Choubey:2009jq,Mena:2014sja,Xu:2014via,Fu:2014isa,Palomares-Ruiz:2015mka,Aartsen:2015ivb,Palladino:2015vna,Arguelles:2015dca,Bustamante:2015waa,Vincent:2016nut}.
The neutrino production mechanisms determine the flavor ratios that leave the source, $f_{\alpha,\text{S}}$ (with $f_{e,\text{S}} + f_{\mu,\text{S}} + f_{\tau,\text{S}} = 1$). If neutrinos are produced in the decay of pions made in proton-photon or proton-proton interactions, then, to first order, $\left( f_{e,\text{S}} : f_{\mu,\text{S}} : f_{\tau,\text{S}} \right) = \left( \frac{1}{3} : \frac{2}{3} : 0 \right)$.

Flavor mixing determines the ratios at Earth: $f_{\alpha,\oplus} = \sum_\beta f_{\beta,\text{S}} P_{\beta\alpha}$.  They depend on the values of the mixing angles, CP-violation phase, and, if decay is present, on energy, lifetimes, and source redshifts.  The source flavor ratios above yield the standard expectation of $\left( f_{e,\oplus} : f_{\mu,\oplus} : f_{\tau,\oplus} \right) \approx \left( \frac{1}{3} : \frac{1}{3} : \frac{1}{3} \right)$.

Decay affects the flavor composition during propagation by depleting the population of heavier mass eigenstates and enhancing the population of the lightest one.  Under complete decay ($D \ll 1$), flavor ratios are given by the flavor content of the sole remaining stable eigenstate\ \cite{Pakvasa:1981ci}, \ie, $f_{\alpha,\oplus} = \left\vert U_{\alpha 1} \right\vert^2$ in the NH and $f_{\alpha,\oplus} = \left\vert U_{\alpha 3} \right\vert^2$ in the IH.  The position and shape of the transition region from no decay to complete decay depend on the lifetimes and fraction of energy given to daughters\ \cite{Mehta:2011qb}.  
For calculational simplicity, in our results below we focus on the simplified case where daughters receive the full parent energy.  \App\ \ref{appendix:FlavorTransition} contains the derivation of flavor ratios at Earth in this case.  This choice does not imply a loss of generality in our treatment and conclusions as long as we lack sufficient data to probe the neutrino spectral shape for a transitional feature from a scenario of no decay to one of complete decay (see \Sec\ \ref{section:SpectralShape}).


\subsection{Managing unidentified sources}

\begin{figure}[t]
 \begin{center}
  \includegraphics[width=0.94\columnwidth, clip=false, trim=0 0.1cm 0 0]{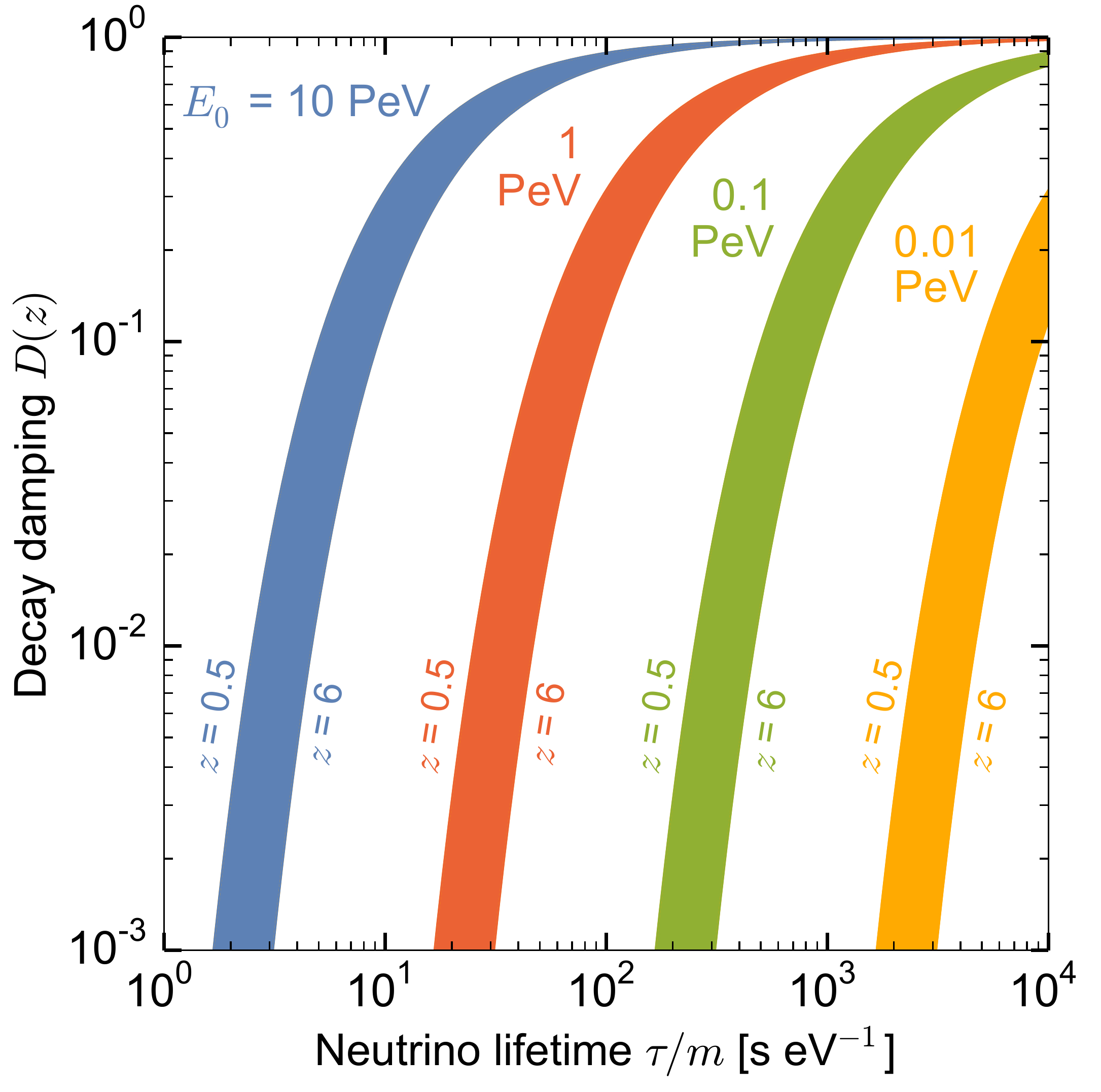}
 \end{center}
 \caption{\label{fig:D-vs-lifetime}Decay damping $D$ as a function of neutrino lifetime, for different values of neutrino energy. The bands are generated by varying the redshift between 0.5 and 6.}
\end{figure}

To compute the decay-induced damping of the neutrino flux emitted by a source, we need to know its redshift; see \equ{DiDefinition}.  With it, we can calculate, for a given neutrino energy, what lifetimes lead to complete decay.  The problem with this specific approach is that no astrophysical neutrino sources have been identified yet\ \cite{Aartsen:2016oji}.

However, we can reasonably assume that the luminosity density of neutrino sources traces the distribution of other sources, such as star-forming regions and active galactic nuclei, both of which peak at $z \approx 1$, or $L \approx 4$ Gpc.  The diffuse neutrino flux at Earth is the added emission from sources at all redshifts.  For a Euclidean universe and no source evolution, the number of sources at distance $L$ rises as $L^2$ while the flux from each falls as $L^{-2}$, meaning that all distances contribute comparably to the total flux; see \App\ \ref{section:DiffuseFlux}.  This is modified by cosmological effects and source evolution.  
\figu{differential-flux-vs-redshift} shows that the dominant contributions to the diffuse flux come from the range $z = 0.5\--1$.  Thus, even though the sources are unidentified, we adequately know their distance.

Serendipitously, in that range, $\mathcal{Z}(z)$ in \equ{DiDefinition} is already close to its asymptotic value, which means that the redshift-dependent part of the damping is naturally nearly as strong as it can be.  This is what allows decay in the diffuse flux to be complete or nearly complete, and what sets our projected sensitivity at 10 s eV$^{-1}$.


\subsection{Managing uncertainties in the energy spectrum}\label{section:SpectralShape}

High-energy astrophysical neutrino data is well-fit by a power-law flux $\Phi\left(E_0\right) = \Phi_0 \left(E_0/\text{100 TeV}\right)^{-\gamma}$\ \cite{Aartsen:2013bka,Aartsen:2013jdh,Aartsen:2013eka,Aartsen:2014gkd,Aartsen:2015knd,Aartsen:2015rwa,Aartsen:2015xup,Aartsen:2016xlq}.  Combining all of the available IceCube data sets --- including events that started inside the detector and through-going muons that crossed it --- yields an all-flavor normalization $\Phi_0 = 6.7_{-1.2}^{+1.1} \cdot 10^{-18}$ GeV$^{-1}$ cm$^{-2}$ s$^{-1}$ sr$^{-1}$ and spectral index $\gamma = 2.50 \pm 0.09$\ \cite{Aartsen:2015knd}.   Using only through-going muons, which reach lower energies, yields a harder spectrum, with a muon-flavor-only normalization $\Phi_0 = 0.90_{-0.27}^{+0.30} \cdot 10^{-18}$ GeV$^{-1}$ cm$^{-2}$ s$^{-1}$ sr$^{-1}$ and $\gamma = 2.13 \pm 0.13$~\cite{Aartsen:2016xlq}.  Above 200 TeV, these fluxes are compatible, assuming flavor equipartition at Earth\ \cite{Aartsen:2016xlq}.  
Below and in \Sec\ \ref{section:DecayShowerRate}, we show that decay effects could be detectable for either value of the spectral index.

\begin{figure}[t!]
 \centering
 \includegraphics[width=\columnwidth]{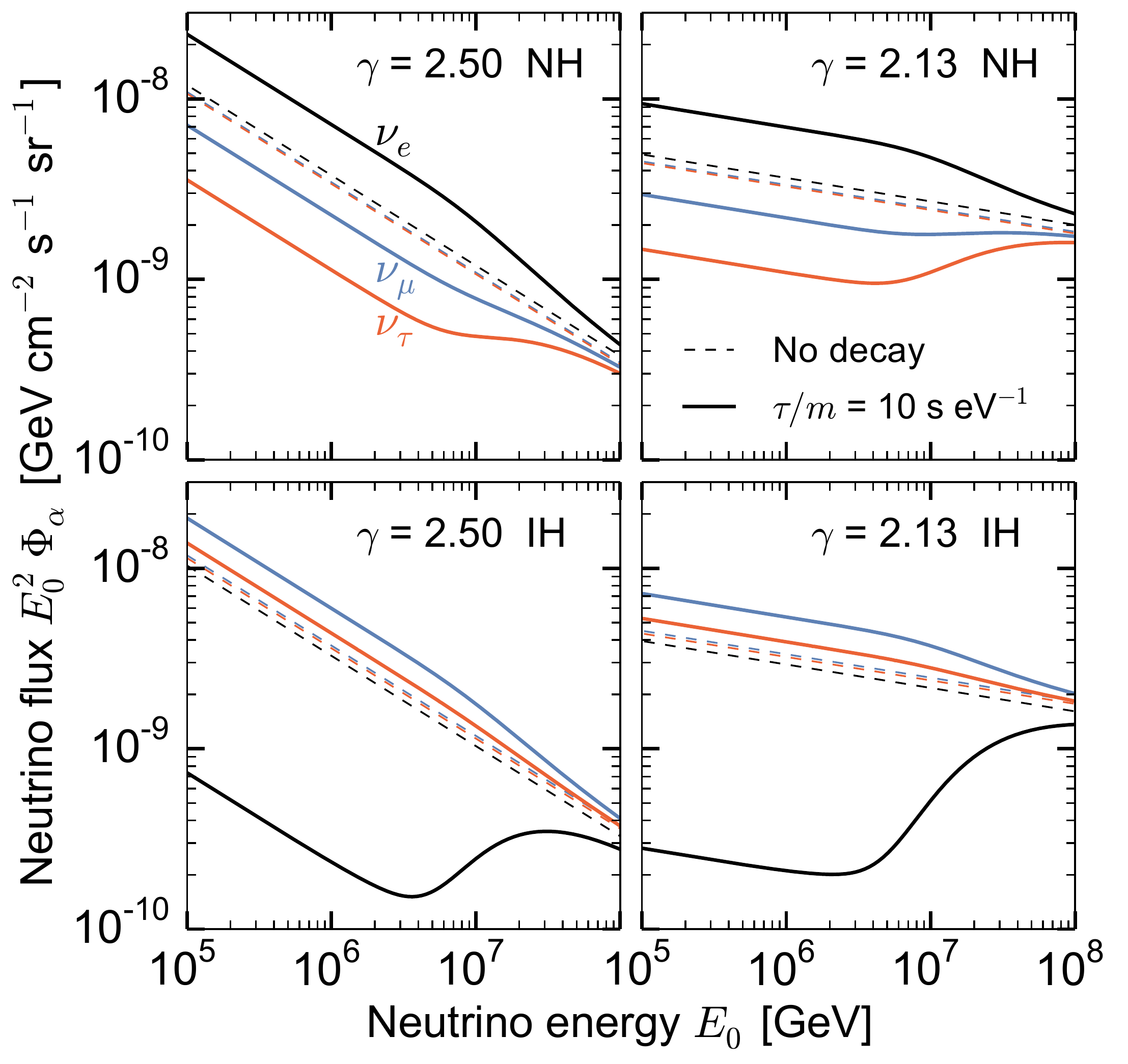}
 \caption{Diffuse neutrino fluxes $\Phi_{\alpha}$ ($\alpha = e,\mu,\tau$) as functions of energy, without and with decay. The fluxes of anti-neutrinos are identical. The flavor ratios at the sources are $\left( \frac{1}{6} : \frac{2}{6} : 0 \right)_\text{S}$ separately for neutrinos and anti-neutrinos.  Mixing parameters are fixed to their best-fit values\ \cite{Gonzalez-Garcia:2014bfa}.  A lifetime of $\tau/m = 10~ \text{s eV}^{-1}$ applies to $\nu_2$, $\nu_3$ (NH), and $\nu_1$, $\nu_2$ (IH).  The ``no decay'' lines for $\nu_\mu$ and $\nu_\tau$ cover each other.  See text for details.}
 \label{fig:fluxes-vs-energy}
\end{figure}

Figure\ \ref{fig:fluxes-vs-energy} shows the diffuse neutrino fluxes at Earth, without and with decay, normalized to the two IceCube analyses; see \App\ \ref{section:DiffuseFlux} for details.
In a fit to data, the normalization would be left as a free parameter, but here we have fixed it for illustration purposes.
For $\gamma = 2.50$, our all-flavor flux is normalized to the IceCube combined-likelihood flux\ \cite{Aartsen:2015knd}.  For $\gamma = 2.13$, our $\nu_\mu + \bar{\nu}_\mu$ flux, without decay, is normalized to the IceCube through-going muon flux\ \cite{Aartsen:2016xlq}; the fluxes of other flavors receive the same normalization, modulated by their flavor ratios.  For our choice of $\tau/m = 10$ s eV$^{-1}$, the transition from no decay to complete decay occurs mostly above the energy range we consider.  Under complete decay in the NH, the flavor ratios equal the flavor content of $\nu_1$ (in the IH, of $\nu_3$).  In \figu{fluxes-vs-energy}, the $\nu_\mu$ and $\nu_\tau$ fluxes are not equal because the best-fit values of $\theta_{23} \neq 45^\circ$, $\delta_\text{CP} \neq 0$, and $\theta_{13} \neq 0$\ \cite{Gonzalez-Garcia:2014bfa}.  In this plot, daughter neutrinos receive the full parent neutrino energy; different energy fractions will affect somewhat the fluxes inside the transition region, but not for complete decay.  While the effects of decay shown here are stark, realities of detection make things more difficult, as described below.

With more data, the assumption of a pure power law could be tested.  If the parent spectrum contains a high-energy cut-off, then active daughters could shift it to lower energies. 
However, unless the cut-off were separately known, it would be hard to test decay this way.

Because of low statistics and because measurements of the diffuse neutrino flux are not sensitive to all flavors equally\ \cite{Aartsen:2015knd}, using only neutrino data to infer both the flavor ratios and the normalization of the flux is challenging.  However, under the assumption that the sources of neutrinos are gamma-ray transparent to either photohadronic or hadronuclear interactions, one can supplement IceCube neutrino measurements with {\it Fermi}-LAT measurements of the gamma-ray background\ \cite{Ackermann:2014usa}, which directly probes the normalization of the neutrino flux\ \cite{Murase:2013rfa,Murase:2015xka}.  If the assumption of decay increased the inferred normalization of the neutrino flux by too much, the associated gamma-ray background would exceed the measurements by {\it Fermi}-LAT, hinting at the effect of decay on the flavor ratios.


\subsection{Managing uncertainties in source flavor ratios}

Since neutrino production mechanisms and conditions are largely unknown, there are large uncertainties in the flavor ratios at the sources.  In spite of them, if decay is complete, it will leave an unmistakable imprint on the flavor ratios at Earth.

The region of allowed flavor ratios at Earth, under standard mixing, is generated by varying flavor ratios at the sources freely and mixing parameters within allowed ranges.  It is surprisingly small.  It was first shown in Fig.~2 of \Ref\ \cite{Bustamante:2015waa} (see also Fig.~1 of \Ref\ \cite{Arguelles:2015dca}); the $3\sigma$ contour is shown here as the ``no decay'' region of \figu{ternary-plot-decay-NH}.  This region and the flavor-content regions of pure $\nu_1$ and pure $\nu_3$ are well-separated, at $> 3 \sigma$. Therefore, barring detection aspects, flavor ratios under standard mixing and under complete decay cannot be confused.

This conclusion holds whether or not different sources emit with different flavor ratios.  It also holds if flavor ratios at the sources vary with energy --- as long as flavor ratios at Earth are measured using events binned in a single, wide energy bin, on account of limited statistics; see the Supplemental Material of \Ref~\cite{Bustamante:2015waa} for details.


\subsection{Summary}

Sources of high-energy astrophysical neutrinos, while undetected, likely trace the redshift distribution of other objects.
Hence, most of the diffuse flux originates from $z \approx 0.5\--1$, which naturally allows decay to have a strong effect.
Additionally, uncertainties in the spectral index of the power-law diffuse flux and in the flavor composition at the sources are unable to mask the effect of decay.


\section{Managing detection aspects}\label{section:ManagingDetection}


\subsection{Flavor measurements in IceCube}\label{subsec:NeutrinoSignatures}

In IceCube, high-energy neutrinos interact with nucleons in the Antarctic ice via deep-inelastic scattering; see \App\ \ref{section:ShowerRate} for details.  The interactions are detected by collecting the Cherenkov light of the final-state particles.

Charged-current interactions create final-state hadrons and charged leptons.  A final-state muon leaves a track of light a few kilometers long that is clearly identifiable.  (Tracks also come from the decay of taus, produced in $\nu_\tau$ interactions, into muons, which occurs 17\% of the time; and, at higher energies, from taus themselves\ \cite{Kistler:2016ask}.)  A final-state electron or tau initiates a localized shower whose light adds to that of the shower initiated by final-state hadrons.  Using the observed energy spectrum of showers allows to identify the astrophysical neutrino component more clearly than using the spectrum of tracks\ \cite{Beacom:2004jb}.  While the particle content of showers created by final-state hadrons, electrons, and taus is different, IceCube is currently insensitive to the difference (muon and neutron echoes might solve this problem\ \cite{Li:2016kra}).  From the relative number of tracks --- mostly from $\nu_\mu$ --- and showers --- mostly from $\nu_e$ and $\nu_\tau$ --- the underlying flavor ratios are inferred.

Neutral-current interactions create final-state hadrons and final-state neutrinos.  Because, on average, hadrons receive a small fraction of the incoming neutrino energy, and because the neutrino spectrum falls with energy, these showers are sub-dominant.

IceCube recently reported the flavor ratios of the diffuse astrophysical neutrino flux\ \cite{Aartsen:2015ivb, Aartsen:2015knd}; their results are shown in Figs.~\ref{fig:ternary-plot-flavor-content-NH} and \ref{fig:ternary-plot-decay-NH}.  They are compatible with the standard expectation of $(\frac{1}{3}:\frac{1}{3}:\frac{1}{3})_\oplus$, as well as with other compositions expected from standard flavor mixing and from various new physics~\cite{Bustamante:2015waa,Arguelles:2015dca}. 

In events that start inside the detector (``high-energy starting events,'' or HESE), the energy of the incoming neutrino can be well reconstructed because all --- for showers --- or a large fraction --- for tracks --- of it is deposited in final-state particles that shower inside the detector.  On the contrary, in through-going track events, the energy of the incoming neutrino must be loosely reconstructed using the relatively short track segment that traverses the detector.  However, this is not a problem for flavor measurements. 
By statistically inferring the $\nu_\mu$ spectrum from the through-going track spectrum,  IceCube has demonstrated that flavor ratios can be inferred from the combined HESE and through-going track data\ \cite{Aartsen:2015knd}, assuming they are constant over a wide enough energy range.  Just like as with standard mixing, under complete decay flavor ratios would be constant and, therefore, the same kind of combined analysis could be used (see, however, the recommendations in \Sec\ \ref{section:Improvements}).

Above $\sim 5$ PeV, flavor-specific detection signatures become accessible\ \cite{Learned:1994wg,Beacom:2003nh,Bugaev:2003sw,Anchordoqui:2004eb,Bhattacharya:2011qu,Bhattacharya:2012fh,Barger:2014iua,Palladino:2015uoa}; none have been observed yet, and low, but observable, event rates are nominally expected.
For $\bar{\nu}_e$ of energies around 6.3 PeV, the Glashow resonance\ \cite{Glashow:1960zz} is expected to increase the shower rate; we will use this to study decay in the IH in \Sec\ \ref{section:DecayShowerRate}.


\subsection{Managing uncertainties in flavor ratios at Earth}

Because muon tracks can be clearly identified, but showers initiated by $\nu_e$ and $\nu_\tau$ cannot presently be distinguished\ \cite{Palomares-Ruiz:2015mka,Palladino:2015zua,Bustamante:2015waa}, the IceCube flavor contours\ \cite{Aartsen:2015ivb,Aartsen:2015knd} in Figs.\ \ref{fig:ternary-plot-flavor-content-NH} and \ref{fig:ternary-plot-decay-NH} are nearly horizontal.  The slight tilt of the contours is due to the smaller average energy deposition of $\nu_\tau$-initiated showers and to the occasional decay of $\nu_\tau$ to $\mu$, which prevents the $\nu_\tau$ fraction from being higher.  The height of the contours is determined by the number of events, while their width is determined by the indistinguishability of $\nu_e$ and $\nu_\tau$.   

In spite of these limitations, \figu{ternary-plot-flavor-content-NH} shows that the flavor-content region of $\nu_1$, expected from complete decay in the NH, is presently disfavored at $\gtrsim 2\sigma$.  This observation is the basis of the method to calculate lifetime sensitivity introduced in \Sec\ \ref{section:DecayFlavorRatios}.   More data would shrink the IceCube flavor contours.  Assuming no other change, this would disfavor more strongly complete decay in the NH; see, {\it e.g.,} \Refs\ \cite{Bustamante:2015waa,Shoemaker:2015qul} for projections using the planned IceCube-Gen2\ \cite{Aartsen:2014njl}.

Progress should move on three fronts.  First, more statistics, gathered either by IceCube or future detectors\ \cite{Katz:2006wv,Aartsen:2014njl,Adrian-Martinez:2016fdl,Avrorin:2016zna}, will reduce mainly the height of the contours.  Second, detection of events at a few PeV may reveal flavor-specific signatures.  The observation of double bangs\ \cite{Learned:1994wg} (or, at lower energies, double pulses\ \cite{Aartsen:2015dlt}) is desirable because it would clearly identify $\nu_\tau$, but it is not essential to test decay.  It would mainly help shape the region of standard allowed flavor ratios (``no decay'' in \figu{ternary-plot-decay-NH}); see Fig.\ 2 in \Ref\ \cite{Shoemaker:2015qul}.  Because this region is roughly aligned with lines of constant $f_{\tau,\oplus}$, improvement would be slight, unless extreme values of $f_{\tau,\oplus}$ are measured or high precision is achieved\ \cite{Li:2016kra}.  On the other hand, the observation of the Glashow resonance\ \cite{Glashow:1960zz}, above $\sim 5$ PeV, would clearly identify $\bar{\nu}_e$ and constitutes a strong test of decay in the IH, as we show in \Sec\ \ref{section:DecayShowerRate}.  Third, breaking the degeneracy between  $\nu_e$- and $\nu_\tau$-initiated showers could reduce the width of the IceCube contours appreciably.  A large improvement in the precision of $\nu_e$ and $\nu_\tau$ flavor ratios could be achieved by detecting muon and neutron echoes\ \cite{Li:2016kra} from showers with energies between 25 TeV and 1 PeV.


\subsection{Need for a clean extragalactic sample}\label{section:Improvements}

To generate the contours of flavor composition in Figs.\ \ref{fig:ternary-plot-flavor-content-NH} and \ref{fig:ternary-plot-decay-NH}, IceCube used all available events with energies between 10 TeV and 2 PeV\ \cite{Aartsen:2015knd}.  However, if flavor composition measurements are to be used to test decay, they must not contain any contamination from non-extragalactic neutrinos.  

For a lifetime of 10 s eV$^{-1}$, there is no decay for atmospheric or even Milky Way neutrinos, because the distances are much less than the Gpc-scale range.  Clearly, if data have a large contamination of such neutrinos, lifetime sensitivities derived from them will be incorrect.

Atmospheric contamination can be averted by restricting the flavor analysis to events with high energies ({\it e.g.}, above 60 TeV\ \cite{Aartsen:2014gkd}).  
Galactic contamination\ \cite{Fox:2013oza,Neronov:2013lza,Razzaque:2013uoa,Ahlers:2013xia,Lunardini:2013gva,Taylor:2014hya,Kachelriess:2014oma,Spurio:2014una,Anchordoqui:2014rca,Gaggero:2015xza,Ahlers:2015moa,Emig:2015dma,Neronov:2015osa,
Palladino:2016zoe,Neronov:2016bnp,Anchordoqui:2016dcp,Halzen:2016seh} can be averted by restricting the flavor analysis to events with high Galactic latitudes.
Events with lower energy and closer to the Galactic Plane should be either discarded or given a reduced significance.

To obtain trustable lifetime limits, dedicated analyses performed by experimental collaborations should implement these restrictions.


\subsection{Summary}

Even though neutrino energy can be reconstructed more accurately with high-energy starting events than with through-going tracks, IceCube has shown that both event types can be combined to infer flavor ratios.   
Flavor measurements, while unable to distinguish between showers initiated by $\nu_e$ and $\nu_\tau$, are already precise enough to disfavor a pure-$\nu_1$ composition, compatible with complete decay in the NH.  Since our proposed analysis hinges on Gpc-scale distances to sources, it must avoid contamination by neutrinos produced closer than that.


\section{Estimating lifetime sensitivities}\label{section:EstimatingLimits}


\subsection{Decay with flavor ratios at present}\label{section:DecayFlavorRatios}

Figure\ \ref{fig:ternary-plot-flavor-content-NH} shows that present IceCube flavor ratios~\cite{Aartsen:2015knd} seemingly already disfavor at $\gtrsim 2\sigma$ complete decay in the NH, \ie, $f_{\alpha,\oplus} = \left\vert U_{\alpha 1} \right\vert^2$, for all values of the mixing parameters within $3\sigma$ (assuming no local contamination). 
Below, we use this observation to estimate the present nominal sensitivity to the lifetimes of $\nu_2$ and $\nu_3$.  We discuss decay in the IH later.

Our nominal sensitivity is set by the values of $\tau_2/m_2$ and $\tau_3/m_3$ for which $f_{\alpha,\oplus} = \left\vert U_{\alpha 1} \right\vert^2$, regardless of uncertainties in the mixing parameters and flavor ratios at the sources.  Since we look for complete decay, we assume, in practice, equal lifetimes, \ie, $\tau_2/m_2 = \tau_3/m_3 \equiv \tau/m$; however, this restriction is not essential.  We proceed by generating regions of allowed flavor ratios for different values of $D$, using \equ{FlavorRatiosEarthDefinition}, and scanning over all possible flavor ratios at the sources and values of the mixing parameters within their $3\sigma$ uncertainties. 

\begin{figure}[t!]
 \centering
 \includegraphics[width=\columnwidth,clip=true,trim=0 0.3cm 0 0.8cm]{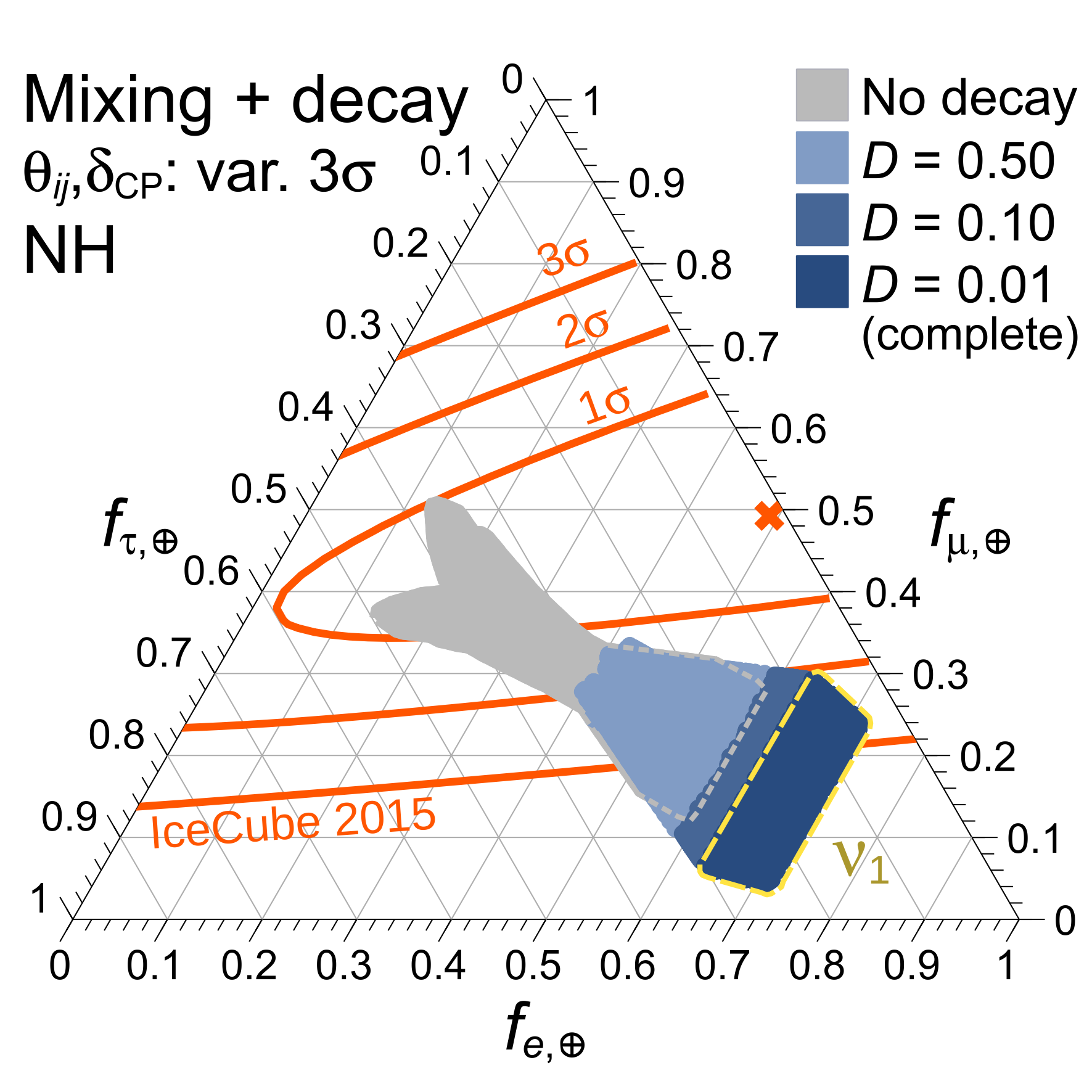}
 \caption{\label{fig:ternary-plot-decay-NH}Allowed $\nu_\alpha + \bar{\nu}_\alpha$ flavor ratios at Earth with decay to $\nu_1$ (NH). For each value of the decay damping $D$, the region is generated by scanning over all possible flavor ratios at the source and mixing parameters within $3\sigma$\ \cite{Gonzalez-Garcia:2014bfa}. The flavor-content region of $\nu_1$ is outlined in dashed yellow\ \cite{Bustamante:2015waa}.}
\end{figure}

Figure\ \ref{fig:ternary-plot-decay-NH} shows the resulting regions.  Decay is complete enough for $D \lesssim 0.01$: the region of allowed flavor ratios is fully contained within the flavor-content region of pure $\nu_1$.  Therefore, $D \lesssim 0.01$ is disfavored at $\gtrsim 2\sigma$.  \figu{D-vs-lifetime} shows that, at energies of $\sim 1~\text{PeV}$, $D = 0.01$ corresponds to a lifetime of $\sim 10~\text{s eV}^{-1}$.  Thus, the nominal IceCube limit achieved with flavor ratios is, roughly, 
\begin{equation}
 \tau_2/m_2, \; \tau_3/m_3 \gtrsim 10~\text{s eV}^{-1} \; (\gtrsim 2\sigma,~\text{NH}) \;.
\end{equation}
This sensitivity is independent of flavor ratios at the sources and $3\sigma$ uncertainties in mixing parameters.  \figu{mass-vs-lifetime}, left panel, shows this is an improvement of $10^4$ and $10^{11}$ over existing limits.  

A more realistic sensitivity calculation should implement the conditions outlined in \Sec\ \ref{section:Improvements}.  These would reduce the number of events and, therefore, widen the flavor contours.  As a result, a realistic lifetime sensitivity could be weaker than our nominal estimate.

\begin{figure*}[t!]
 \begin{center}
  \includegraphics[width=0.985\columnwidth]{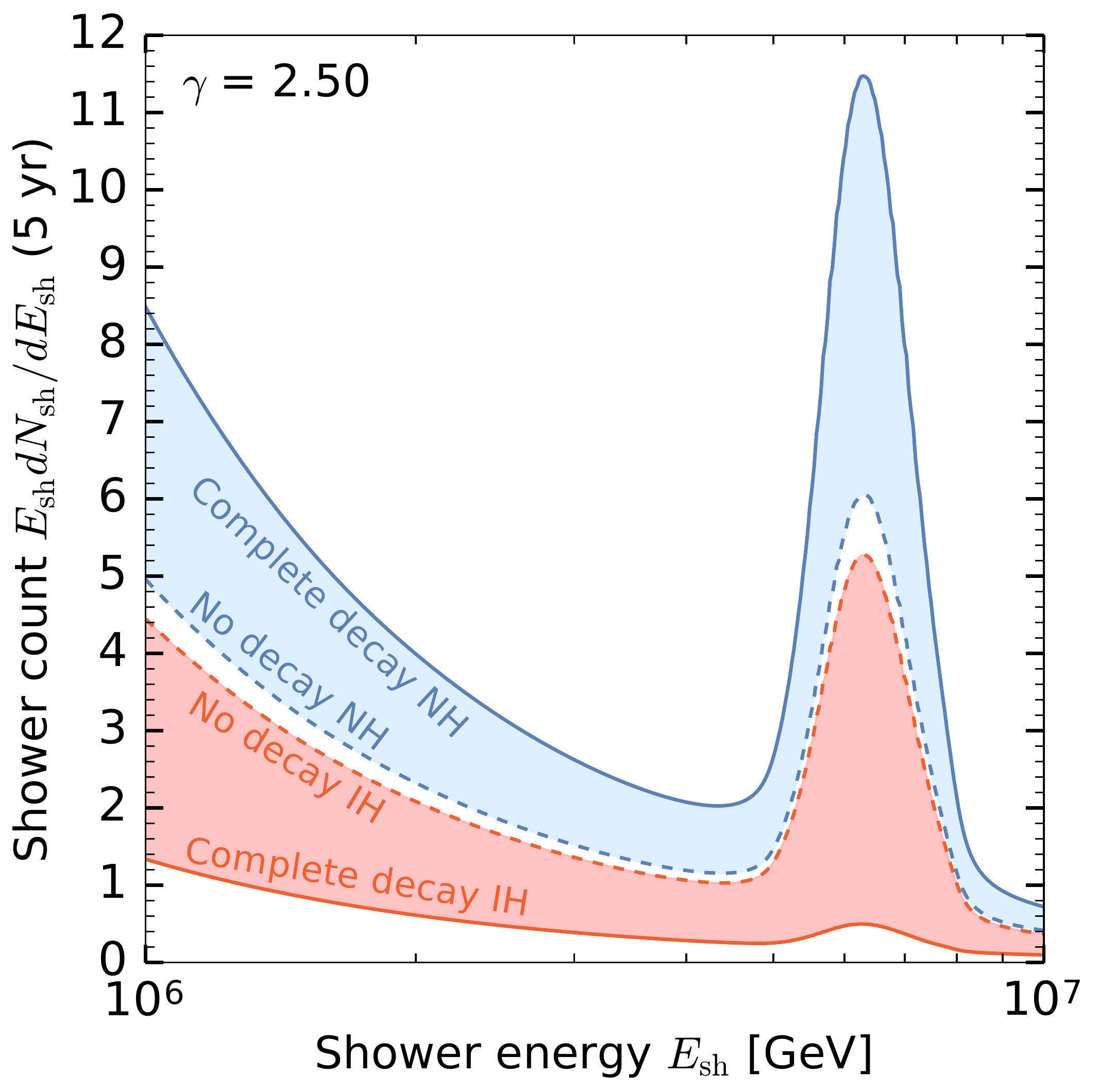}
  \includegraphics[width=0.985\columnwidth]{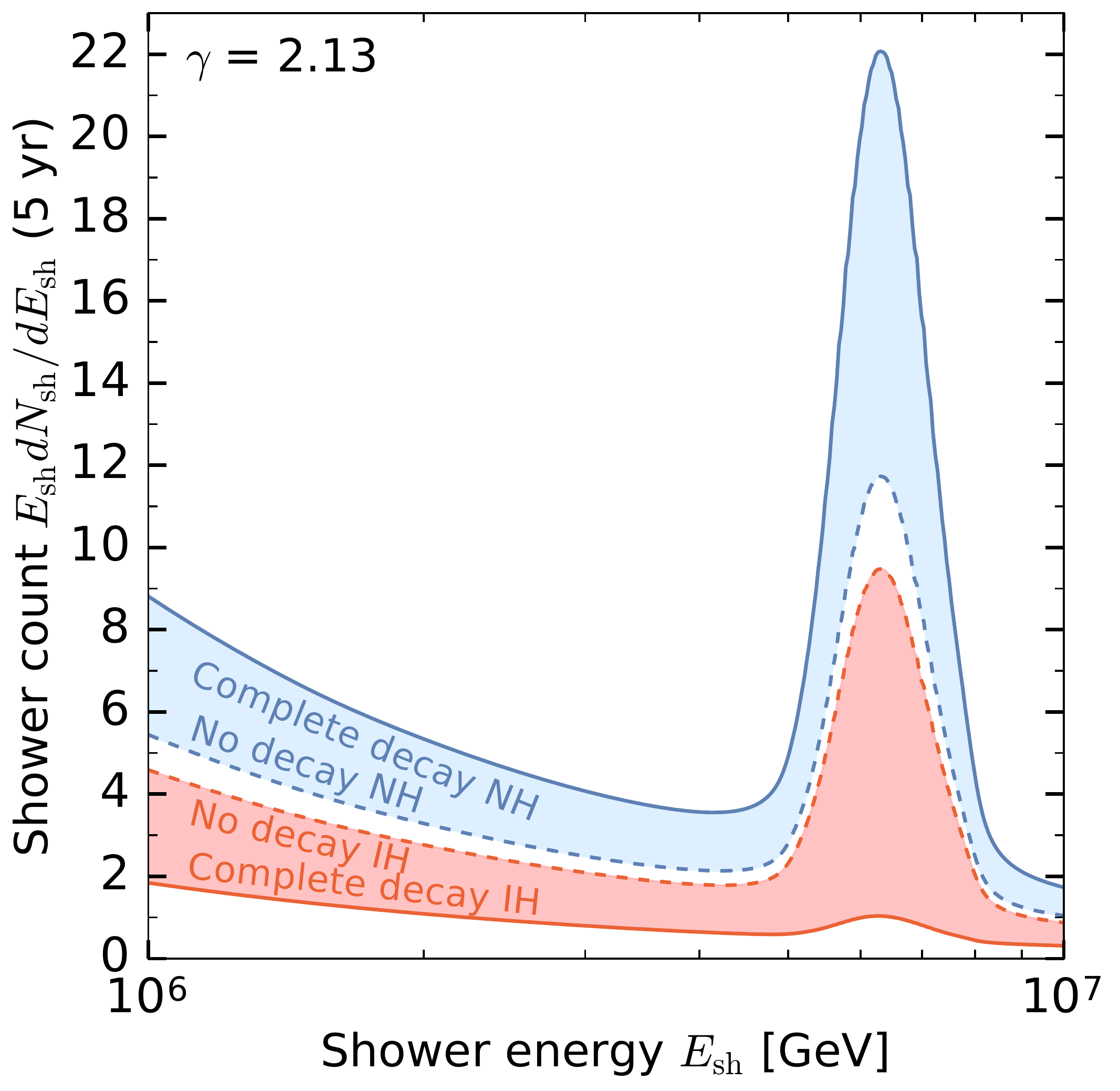}
 \end{center}
 \caption{\label{fig:cascade-rates}Shower spectrum at IceCube, assuming five years of exposure.  The detector energy resolution is set to $\delta E_\text{sh} / E_\text{sh} = 0.1$\ \cite{Aartsen:2013vja}. {\it Left:} Using a flux $\propto E^{-2.50}$\ \cite{Aartsen:2015knd}. {\it Right:} Using a flux $\propto E^{-2.13}$\ \cite{Aartsen:2016xlq}.  Note the change in scale.  Contributions of $\nu_\tau$-initiated showers are not added.  See text for details.}
\end{figure*}

Presently, we cannot use measured flavor ratios to strongly limit complete decay in the IH, because $f_{\alpha,\oplus} = \left\vert U_{\alpha 3} \right\vert^2$ is only weakly disfavored, between $1\sigma$ and $2\sigma$; see \figu{ternary-plot-flavor-content-NH}.  However, more neutrino data might not only tighten the contours, but also shift them up or down.  A higher ratio of tracks to showers would shift the curves up, towards higher values of $f_{\mu,\oplus}$.  So would a lower fraction of tracks that are mis-classified as showers; currently, this is about $30\%$~\cite{Aartsen:2015ivb}.  Complete decay in the NH would be even more disfavored and complete decay in the IH would be more compatible with the data.  Conversely, a lower ratio of tracks to showers would allow complete decay in the NH and would disfavor complete decay in the IH.  In the latter case, we could use the method outlined above to find lifetime limits in the IH.

\Ref\ \cite{Pagliaroli:2015rca} found a $2\sigma$ exclusion of decay in the IH using IceCube data.  However, it did so by using the tracks-to-showers ratio of reported events, whereas here we have directly used the flavor contours provided by IceCube, which contain all of the detector systematics from their combined maximum-likelihood analysis\ \cite{Aartsen:2015knd}.


\subsection{Decay in the high-energy shower rate}\label{section:DecayShowerRate}

We can also test decay using high-energy showers.  This works for both hierarchies, but we focus on the IH because it gives the cleanest signal and because it is not presently strongly constrained by flavor ratios.

Out of the nine flavor fractions of the three mass eigenstates, the electron-flavor content of $\nu_3$ is unique: it is, by far, the smallest one.  In particular, it is an order or magnitude smaller than the electron-flavor content of $\nu_1$ and $\nu_2$, \ie, $\lvert U_{e 3} \rvert^2 \approx 0.03$ versus $\lvert U_{e 1} \rvert^2 \approx 0.67$ and $\lvert U_{e 2} \rvert^2 \approx 0.30$.  Therefore, an observable that is highly sensitive to $f_{e,\oplus}$ could distinguish between a flux that contains comparable proportions of all eigenstates --- as expected from standard mixing --- and a flux that contains exclusively $\nu_3$ --- as expected from complete decay in the IH.  The rate of high-energy showers, driven by the Glashow resonance, is such an observable.

For shower energies above $\sim 5$ PeV, the Glashow resonance\ \cite{Glashow:1960zz} provides a way to break the degeneracy between showers initiated by $\nu_e$ and $\nu_\tau$\ \cite{Bhattacharya:2011qu,Bhattacharya:2012fh,Barger:2014iua,Palladino:2015uoa}.  Only $\bar{\nu}_e$ with energies around 6.3 PeV trigger the resonant process $\bar{\nu}_e + e \to W^-$.  The on-shell $W$ decays hadronically 67\% of the time\ \cite{Agashe:2014kda}, increasing the shower rate.  Therefore, the rate of showers around the resonance energy is a direct probe of the $\bar{\nu}_e$ content.

To estimate the rate of high-energy showers in IceCube, we extrapolate the power-law astrophysical diffuse flux to energies above 2 PeV.  Even with the Glashow resonance, the expected rate is low\ \cite{Palladino:2015uoa}.   (Non-hadronic decays of the $W$ offer alternative detection signals, but with even lower rates\ \cite{Bhattacharya:2011qu,Bhattacharya:2012fh}.)  Even so, decay signatures are stark, especially in the IH, and could be detectable.

Since the enhancement from Glashow resonance depends on $\bar{\nu}_e$, in what follows we separate the flavor ratios of neutrinos ($f_\alpha$) and anti-neutrinos ($f_{\bar{\alpha}}$).  For illustration, we restrict the discussion to equal neutrino and anti-neutrino flavor ratios at the sources: $\left( f_{e,\text{S}} : f_{\mu,\text{S}} : f_{\tau,\text{S}} \right) = \left( f_{\bar{e},\text{S}} : f_{\bar{\mu},\text{S}} : f_{\bar{\tau},\text{S}} \right) = \left( \frac{1}{6} : \frac{2}{6} : 0 \right)$, where the new normalization condition is $\sum_\alpha f_{\alpha,\text{S}} + f_{\bar{\alpha},\text{S}} = 1$ (see \Refs\ \cite{Nunokawa:2016pop,Vincent:2016nut} for the case of different flavor ratios of $\nu$ and $\bar{\nu}$).  This composition is expected, to first order, from neutrino production via $pp$ interactions.  (To first order, in $p\gamma$ interactions no $\bar{\nu}_e$ are produced unless the target photons are thermal\ \cite{Kelner:2008ke,Hummer:2010vx} (\eg, in choked jets\ \cite{Murase:2013ffa}, quasar-hosted blazars\ \cite{Murase:2014foa}, pulsars\ \cite{Murase:2009pg}, \etc), in which case the ratio of neutrinos to anti-neutrinos can be $\sim$ 1.  We comment on this below.)

Figure\ \ref{fig:cascade-rates} shows the shower spectra $E_\text{sh} dN/dE_\text{sh}$ for the two IceCube fluxes introduced in \Sec\ \ref{section:SpectralShape}.  We do not include the contribution from $\nu_\tau$ and $\bar{\nu}_\tau$ charged-current interactions because we expect that, at these energies, they should be separately identifiable as double pulses\ \cite{Aartsen:2015dlt}, double bangs\ \cite{Learned:1994wg}, lollipops\ \cite{Beacom:2003nh}, or tau-to-muon decays \cite{DeYoung:2006fg}.  See \App\ \ref{section:ShowerRate} for details.  We have fixed the values of the mixing parameters at their best-fit values\ \cite{Gonzalez-Garcia:2014bfa}.  The number of events in an energy range can be estimated by multiplying the height of the curve times $2.3 \cdot \Delta \log_{10} ( E_\text{sh} )$.  At these energies, conventional atmospheric neutrinos contribute negligibly\ \cite{Laha:2013lka}, but prompt neutrinos could be relevant\ \cite{Laha:2016dri}.  Shower rates for $\gamma = 2.13$ are roughly twice as large as for $\gamma = 2.50$.

\begin{figure*}[t!]
 \begin{center}
  \includegraphics[width=1.025\columnwidth]{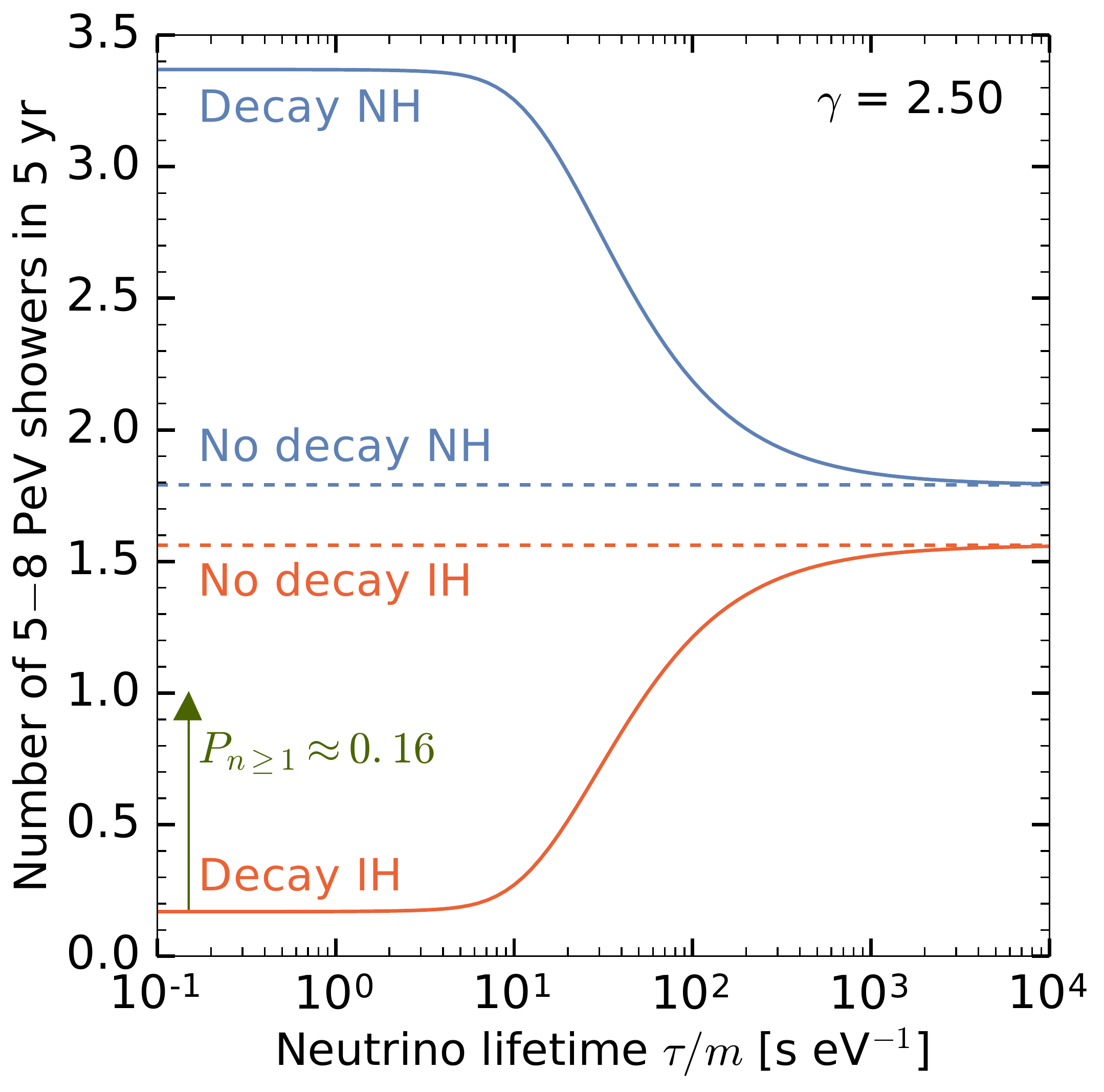}
  \includegraphics[width=0.99\columnwidth]{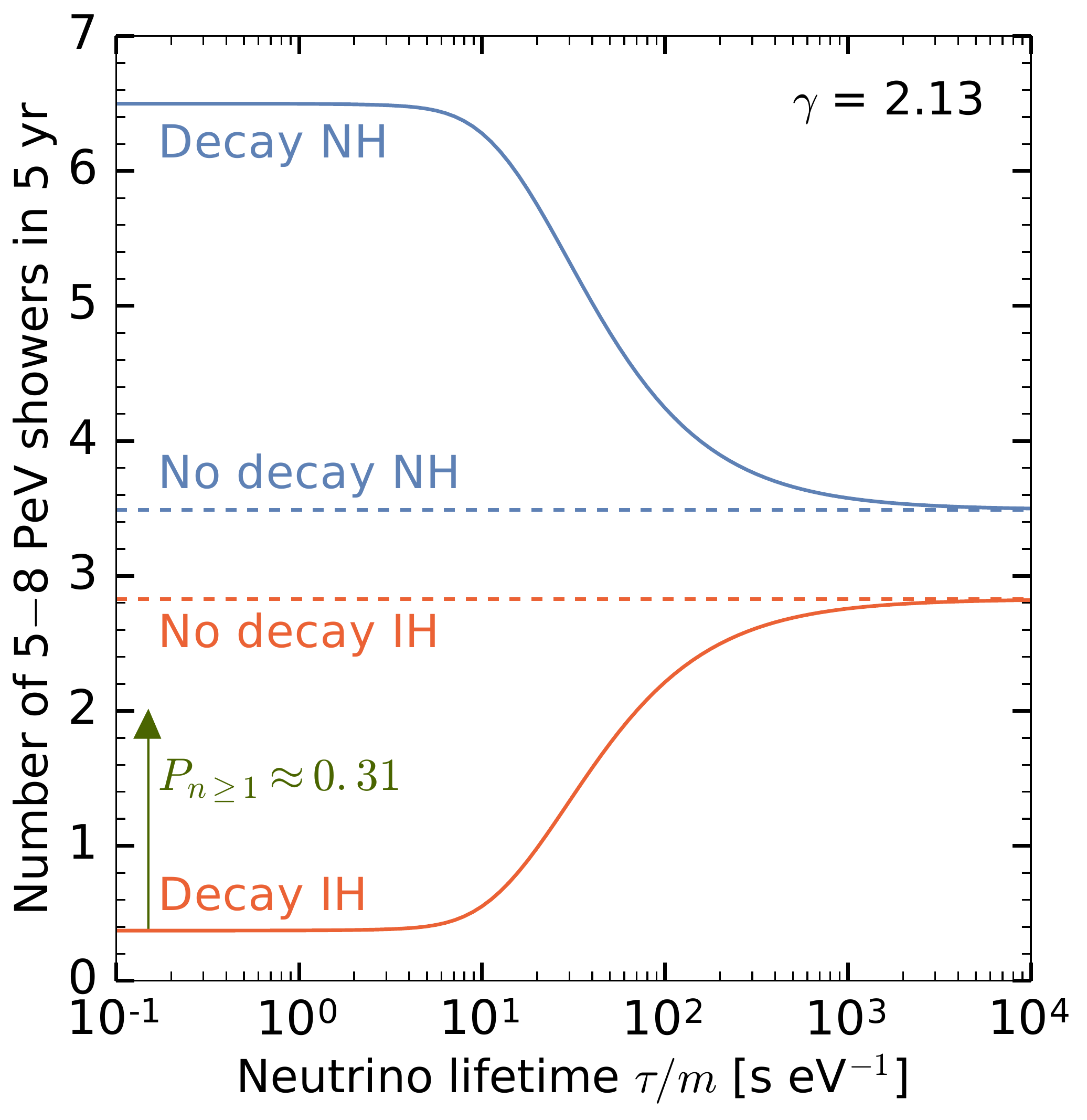}
 \end{center}
 \caption{\label{fig:cascades-vs-lifetime}Number of showers in IceCube in the range 5--8 PeV, as a function of the common lifetime of the two heavier mass eigenstates, assuming five years of exposure.  {\it Left:} Using a flux $\propto E^{-2.50}$\ \cite{Aartsen:2015knd}. {\it Right:} Using a flux $\propto E^{-2.13}$\ \cite{Aartsen:2016xlq}.  Note the change in scale.  The probability $P_{n \geq 1}$ of detecting one or more events under complete decay in the IH is only $\sim 16\%$ (left) or $\sim 31\%$ (right).  Therefore, if even a single event is detected in the energy range of the Glashow resonance, that will disfavor complete decay in the IH.  With higher statistics, the significance will increase rapidly.  See text for details.}
\end{figure*}

Figure\ \ref{fig:cascades-vs-lifetime} shows the integrated number of showers in the range 5--8 PeV, which brackets the Glashow resonance energy, as a function of the lifetime of the two heavier mass eigenstates.  Decay is complete for $\tau/m \lesssim 10$ s eV$^{-1}$, and unobservable for $\tau/m \gtrsim 100$ s eV$^{-1}$.  The Glashow resonance yield completely dominates the continuum yield in this energy range: without decay, the continuum accounts for less than 20\% of events.  Hence, the first shower detected in that range will mean a detection of the resonance, provided that the continuum at lower energies is measured in a way that suggests its continuation to the 5--8 PeV range (and assuming no harder spectral component hides at these energies or close above it).  This means high-energy showers do probe just the $\bar{\nu}_e$ fraction.

Under complete decay in the NH, the integrated shower rate is enhanced by a factor of $\lvert U_{e 1} \rvert^2 / (1/3) \approx 2$.  For $\gamma = 2.50$, it changes from 1.8 to 3.4 events in five years; for $\gamma = 2.13$, it changes from 3.5 to 6.5 events.  In both cases, the change may be difficult to distinguish.

Under complete decay in the IH, the integrated shower rate is depleted by a factor of $\lvert U_{e 3} \rvert^2 / (1/3) \approx 0.1$.  The average shower rate becomes small: for $\gamma = 2.50$, the rate is less than 0.2 events in five years, so the probability of observing 1 or more events is $\sim 16 \%$; for $\gamma = 2.13$, the rate is roughly twice that, so the probability is $\sim 31\%$.   This makes the prediction of small shower rates under complete decay in the IH relatively robust.

By itself, the non-detection of high-energy showers in five years cannot be unequivocally attributed to complete decay in the IH --- this could be equally due to a cut-off below the Glashow resonance energy\ \cite{Aartsen:2015knd}, neutrino production via $p\gamma$, or decay. 

{\it However, the detection of one event would disfavor complete decay in the IH and, therefore, could be used to set lifetime limits.}  This corresponds to a nominal sensitivity of
\begin{equation}
 \tau_1/m_1, \; \tau_2/m_2 \gtrsim 10\ \text{s eV}^{-1}  \; (\sim 2\sigma,~\text{IH}) \;,
\end{equation}
for $\gamma = 2.50$ ($\sim$ $1\sigma$ for $\gamma = 2.13$).  Figure\ \ref{fig:mass-vs-lifetime}, right panel, shows this is an improvement of $10^4$ over existing limits.  
The significance we quote is that of the low event rate at complete decay fluctuating up to yield one event in five years.  While the significance is lower for $\gamma = 2.13$, this is offset by a higher expected number of events.

When the statistics get higher, say, with IceCube-Gen2, the details of the argument would change, but would still rely on the dominance of the Glashow resonance over the underlying continuum.  Detection of two events would rule out complete decay in the IH at $\sim 5\sigma$ ($\sim 3\sigma$ for $\gamma = 2.13$).  IceCube-Gen2 might have an effective area six times larger\ \cite{Aartsen:2014njl}; for the same exposure time, and depending on the spectral index, this would lead to $\sim$ 9--17 events without decay, versus $\sim$ 1--2 events at complete decay in the IH, providing a clearer signal at $\gtrsim 5\sigma$.

\vspace*{-0.3cm}

With more statistics, combining flavor ratios and spectral information could yield stronger limits and reveal the transition to complete decay.  By doing this, \Ref~\cite{Shoemaker:2015qul} estimated that IceCube-Gen2 could reach $\tau/m \gtrsim 500$ s eV$^{-1}$ in ten years, depending on the spectral index.


\section{Summary and conclusions}\label{section:SummaryConclusions}

We have shown that the diffuse flux of high-energy astrophysical neutrinos recently discovered by IceCube can be used to robustly test decay.
Improved limits on neutrino masses have opened up the possibility of a hierarchical mass scheme, allowing for a more model-independent exploration of decay.
We have shown that, in spite of uncertainties in neutrino properties, source properties, and detection aspects, clear tests of decay are possible.

We have provided a roadmap for how dedicated decay analyses should be performed.  For illustration, we have estimated the order-of-magnitude sensitivity of IceCube to neutrino lifetime, using present data and near-future prospects.

First, we have used the flavor composition at Earth of the diffuse flux. In the extreme case of complete decay, all unstable neutrino mass eigenstates decay en route, so the flavor composition of the flux is that of the single remaining, lightest eigenstate, which we assume to be stable.  In the normal mass hierarchy (NH), this is $\nu_1$; in the inverted hierarchy (IH), it is $\nu_3$.   We have shown that present flavor measurements by IceCube seemingly disfavor complete decay in the NH at $\gtrsim 2\sigma$, regardless of flavor composition at the sources and values of mixing parameters.  This translates into a sensitivity to the lifetimes of $\nu_2$ and $\nu_3$ of $\tau / m \gtrsim 10~\text{s eV}^{-1}$, an improvement of $10^4$ and $10^{11}$, respectively, over existing limits. 

Second, we have used the potential near-future detection of high-energy (5--8 PeV) showers in IceCube to probe complete decay in the IH.  Without decay, the shower rate is enhanced by the Glashow resonance, centered at 6.3 PeV.  In contrast, complete decay would make the rate small.  Therefore, the observation of even a single shower in five years would set a lower limit on the lifetimes of $\nu_1$ and $\nu_2$ of 10 s eV$^{-1}$, also at $\gtrsim 2\sigma$ (for a spectral index of $2.50$), an improvement of $10^4$.  With higher statistics, collected either with IceCube or IceCube-Gen2, the significance will increase rapidly.

The observability of decay hinges on the Gpc-scale distances to sources.  To reduce contamination from neutrinos produced too close, we advocate performing dedicated analyses that disfavor any possible events from the atmosphere or the Milky Way.  With more statistics, reduced neutrino and source uncertainties, and improved detection techniques, the sensitivity could be greatly improved.

The new mediator driving neutrino decay could also induce new neutrino-neutrino interactions, which could affect early cosmic history.  Further, IceCube could see the effects of interactions between PeV neutrinos and the cosmological neutrino background as distortions of the power-law spectrum (see, \eg, \Refs\ \cite{Ioka:2014kca,Ng:2014pca,Blum:2014ewa}). The non-detection of these features (so far) puts bounds on the new couplings.  Assuming this mediator is the same one that drives neutrino decay, then these bounds would also be bounds on the neutrino lifetime.  However, exploring these effects lies beyond the scope of this paper. 

If decay is ruled out with astrophysical neutrinos, then searches for new physics with solar, atmospheric, and terrestrial neutrinos will have to be more focused, having fewer possibilities to explain any deviations from standard expectations.  Conversely, if hints of decay are found, that would be important to take into account for cosmological tests of neutrino mass.


\section*{Acknowledgements}

We thank Markus Ahlers, Carlos Arg\"uelles, Kfir Blum, Ranjan Laha, Shirley Li, Tim Linden, Kenny Ng, Sandip Pakvasa, Andrea Palladino, Ian Shoemaker, Aaron Vincent, Walter Winter, and, especially, Sergio Palomares-Ruiz for useful discussion and comments.  JFB and MB are supported by NSF Grant PHY-1404311.  KM is supported by NSF Grant PHY-1620777. MB and KM thank the Institute for Nuclear Theory at the University of Washington for its hospitality during June 2015, and the Department of Energy for partial support during the development of this work.



%

\appendix

\onecolumngrid


\section{Derivation of the flavor-transition probability including neutrino decays}\label{appendix:FlavorTransition}

For concreteness, let us first assume a normal hierarchy (NH), where $\nu_1$ is the lightest and sole stable neutrino mass eigenstate, and the two other active eigenstates decay to it, \ie, $\nu_{2,3} \to \nu_1 + X$, with $X$ any additional decay products that are undetected. Due to the rapid oscillations of the flavor-transition probability, flavor oscillations average out soon after emission; astrophysical neutrinos propagate as an incoherent mix of mass eigenstates (see, however, \Ref\ \cite{Lindner:2001fx}, where decay and oscillation are jointly considered). In the absence of decays, the probability for $\nu_\alpha \to \nu_\beta$ ($\alpha, \beta = e,\mu,\tau$) has the well-known expression $P_{\alpha\beta} = \sum_{i=1}^3 \left\vert U_{\alpha i} \right\vert^2 \left\vert U_{\beta i} \right\vert^2$, dependent only on the components of the lepton mixing matrix, and independent of neutrino energy.

In the presence of decay, we need to consider separately the initial number of mass eigenstate $\nu_i$ at the source, $\hat{N}_i$, and the number that arrives at Earth, $N_i$. Consider briefly decays of the type $\nu_i \to X$, into ``invisible'' products only, \ie, products that are undetected by neutrino experiments.  \Ref~\cite{Baerwald:2012kc} found that the probability in this case is
\begin{equation}
 P_{\alpha\beta}^\text{inv}\left(E_0,z\right) = \sum_{i=1}^3 \left\vert U_{\alpha i} \right\vert^2 \left\vert U_{\beta i} \right\vert^2 \frac{N_i\left(E_0,z,\tau_i/m_i\right)}{\hat{N}_i} \;,
\end{equation}
where the ratio $D_i\left(E_0,z\right) \equiv D\left(E_0,z,\tau_i/m_i\right) \equiv N_i\left(E_0,z,\tau_i/m_i\right) / \hat{N}_i $, shown in \equ{DiDefinition}, is the solution of the redshift-dependent decay equation.  Via decay, the probability has picked up a dependence on the redshift of the source $z$, the received energy of the neutrino $E_0$, and the lifetimes $\tau_i/m_i$ of the mass eigenstates ($\tau_i/m_i \to \infty$ if $\nu_i$ is stable). 

In decays into visible products, however, it is necessary to modify this expression to account for the fact that the decays of the two heavier eigenstates contribute to the flux of the stable one. For the NH, the probability is
\begin{equation}\label{equ:ProbDecayNHIntermediate}
 P_{\alpha\beta}^\text{vis,NH} =
 \left\vert U_{\alpha 1} \right\vert^2 \left\vert U_{\beta 1} \right\vert^2 \left[ \frac{N_1+\left(\hat{N}_2-N_2\right)+\left(\hat{N}_3-N_3\right)}{\hat{N}_1}  \right]
 + \left\vert U_{\alpha 2} \right\vert^2 \left\vert U_{\beta 2} \right\vert^2 \frac{N_2}{\hat{N}_2}
 + \left\vert U_{\alpha 3} \right\vert^2 \left\vert U_{\beta 3} \right\vert^2 \frac{N_3}{\hat{N}_3} \;.
\end{equation}
Here, $\hat{N}_i-N_i$ is the number of $\nu_i$ that remain at detection time. A more useful expression is $( \hat{N}_i-N_i ) / \hat{N}_1 = ( \hat{N}_i/\hat{N}_1 ) ( 1 - D_i )$. The number of $\nu_i$ emitted by the source is a fraction of the total number of neutrinos emitted $\hat{N}_\text{tot}$, namely, $\hat{N}_i = f_{i,\text{S}} \hat{N}_\text{tot}$. With this, the ratio $\hat{N}_i/\hat{N}_1$ is simply the ratio of mass eigenstate flavor ratios, $f_{i,\text{S}}/f_{1,\text{S}}$. Typically, the flavor ratios $\left( f_{e,\text{S}} : f_{\mu,\text{S}} : f_{\tau,\text{S}} \right)$, not the mass eigenstate ratios, are given. The latter can be computed from the former as $f_{i,\text{S}} = \sum_\alpha f_{\alpha,\text{S}} \left\vert U_{\alpha i} \right\vert^2$. 

Thus, for given flavor ratios at the source --- or, equivalently, for given mass-eigenstate ratios at the source --- we can rewrite the flavor-transition probability in the NH, \equ{ProbDecayNHIntermediate}, as
\begin{eqnarray}\label{equ:ProbabilityNH}
 P_{\alpha\beta}^\text{vis,NH}\left(E_0,z\right) &=& 
 \left\vert U_{\alpha 1} \right\vert^2 \left\vert U_{\beta 1} \right\vert^2 \left\{ D_1\left(E_0,z\right)
 + \frac{f_{2,\text{S}}}{f_{1,\text{S}}} \left[ 1 - D_2\left(E_0,z\right) \right] 
 + \frac{f_{3,\text{S}}}{f_{1,\text{S}}} \left[ 1 - D_3\left(E_0,z\right) \right] \right\} \nonumber \\
 && + \left\vert U_{\alpha 2} \right\vert^2 \left\vert U_{\beta 2} \right\vert^2 D_2\left(E_0,z\right)
 + \left\vert U_{\alpha 3} \right\vert^2 \left\vert U_{\beta 3} \right\vert^2 D_3\left(E_0,z\right) \;.
\end{eqnarray}
Similarly, for the inverse hierarchy (IH), we can write
\begin{eqnarray}\label{equ:ProbabilityIH}
 P_{\alpha\beta}^\text{vis,IH}\left(E_0,z\right) &=& 
 \left\vert U_{\alpha 1} \right\vert^2 \left\vert U_{\beta 1} \right\vert^2 D_1\left(E_0,z\right)
 + \left\vert U_{\alpha 2} \right\vert^2 \left\vert U_{\beta 2} \right\vert^2 D_2\left(E_0,z\right) \nonumber \\
 && + \left\vert U_{\alpha 3} \right\vert^2 \left\vert U_{\beta 3} \right\vert^2 \left\{ 
 \frac{f_{1,\text{S}}}{f_{3,\text{S}}} \left[ 1 - D_1\left(E_0,z\right) \right]
 + \frac{f_{2,\text{S}}}{f_{3,\text{S}}} \left[ 1 - D_2\left(E_0,z\right) \right] 
 + D_3\left(E_0,z\right) \right\} \;.
\end{eqnarray}

Eqs.~(\ref{equ:ProbabilityNH}) and (\ref{equ:ProbabilityIH}) can be combined into a single expression, \ie,
\begin{equation}\label{equ:ProbabilityFull}
 P_{\alpha\beta}\left(E_0,z\right)
 = \left\vert U_{\alpha l} \right\vert^2 \left\vert U_{\beta l} \right\vert^2
   \left\{ 1 + \sum_{j \neq l} \frac{f_{j,\text{S}}}{f_{l,\text{S}}} \left[ 1- D\left(E_0,z,\tau_j/m_j\right) \right] \right\}
   + \sum_{j \neq l} \left\vert U_{\alpha j} \right\vert^2 \left\vert U_{\beta j} \right\vert^2 D\left(E_0,z,\tau_j/m_j\right) \;,
\end{equation}
where $\nu_1$ is stable in the NH ($l = 1$) and $\nu_3$ is stable in the IH ($l = 3$).  There is an important implicit assumption in this derivation: {\it the daughter neutrino receives the full energy of the parent}.  The formalism is not valid otherwise.  A more general treatment will be presented elsewhere.

Formally, these expressions are no longer probabilities, since they can have values greater than one. We maintain the notation $P_{\alpha\beta}$, but \equ{ProbabilityFull} should be understood rather as a flux-modifying factor, not as a flavor-transition probability.

The flavor ratios at Earth $f_{\alpha,\oplus} = \sum_\beta P_{\beta\alpha} f_{\beta,\text{S}}$ can be written as
\begin{equation}\label{equ:FlavorRatiosEarthDefinition}
 f_{\alpha,\oplus} =
 \lvert U_{\alpha l} \rvert^2
 + \sum_{j \neq l}
 f_{j,\text{S}}
 \left( 
 \lvert U_{\alpha j} \rvert^2 - \lvert U_{\alpha l} \rvert^2
 \right)
 D\left(E_0,z,\tau_j/m_j\right) \;.
\end{equation}


\section{The diffuse flux with decay}\label{section:DiffuseFlux}

\setcounter{figure}{0}
\renewcommand{\thefigure}{B\arabic{figure}}
\begin{figure}[t!]
 \centering
 \includegraphics[width=0.5\textwidth]{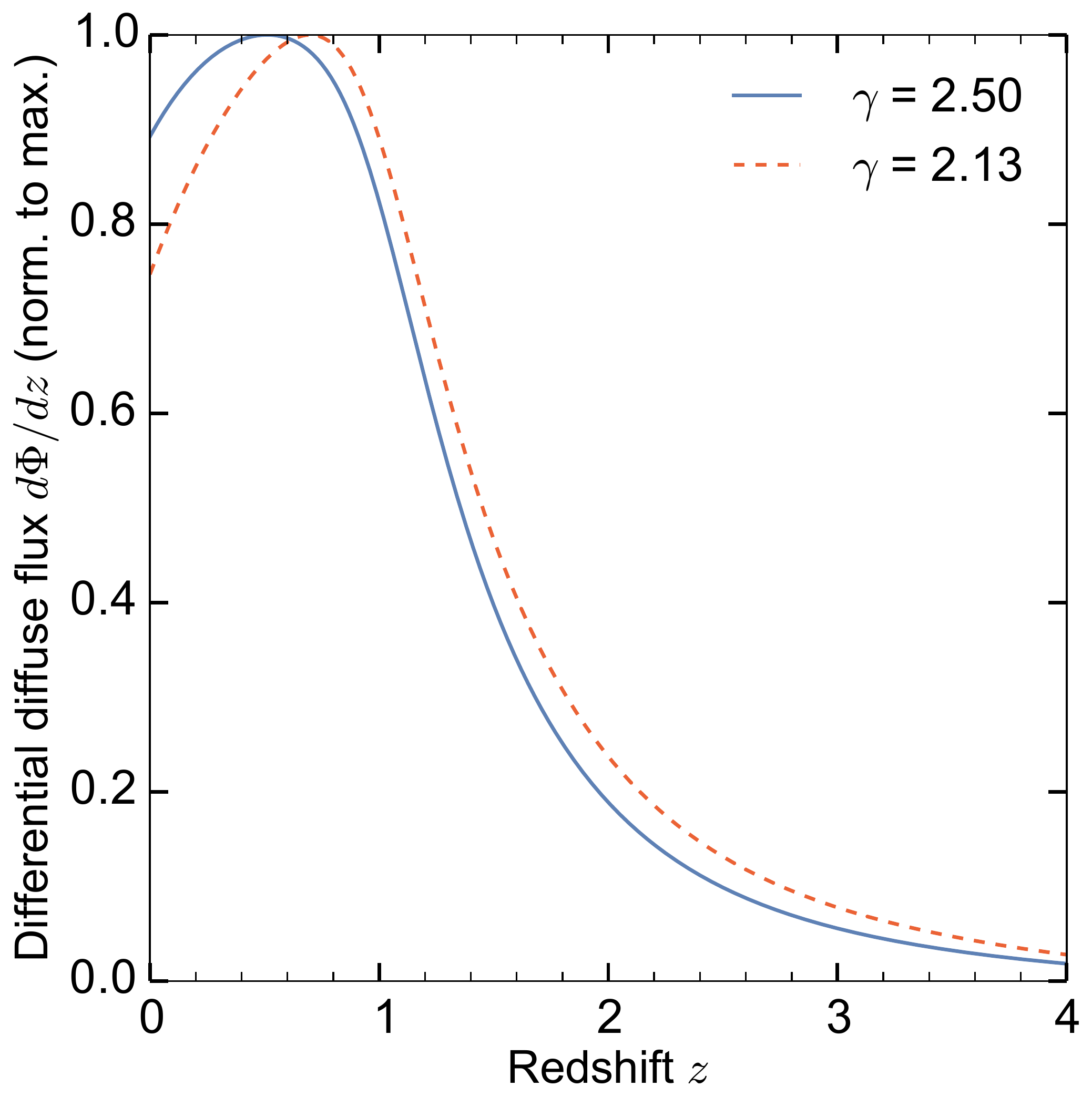}
 \caption{Differential neutrino diffuse flux $d\Phi/dz$, from \equ{DiffuseFlux3}, as a function of redshift, for two choices of spectral index $\gamma$, following IceCube results.  Each curve is individually normalized to its maximum value.}
 \label{fig:differential-flux-vs-redshift}
\end{figure}

We assume that all sources emit the same luminosity $J_{\nu_\beta}\hspace{-0.1cm}\left(E\right)$ (in units of GeV s$^{-1}$) of $\nu_\beta$ ($\beta = e,\mu,\tau$). (For anti-neutrinos, all the expressions below hold, with $\nu \to \bar{\nu}$, $\Phi_\alpha \to \Phi_{\bar{\alpha}}$, and $f_{\alpha,\oplus} \to f_{\bar{\alpha},\oplus}$.)  For the all-flavor luminosity, we assume a power law $J_{\nu_\text{all}}\left(E\right) \propto E^{2-\gamma}$.  The spectral index is fixed either at $\gamma = 2.50$ or $2.13$ in the main text.  Hence, the diffuse energy flux of $\nu_\alpha$ at Earth (in units of GeV cm$^{-2}$ s$^{-1}$ sr$^{-1}$) is (see, {\it e.g.}, \Ref\ \cite{Iocco:2007td})
\begin{equation}
 E_0^2 \Phi_{\alpha}\hspace{-0.1cm}\left(E_0\right)
 =
 \int_0^\infty dz \, 
 \frac{ \rho_\text{src} \left(z\right) } { 4 \pi r^2 \left(z\right) } 
 \cdot 
 \frac{dV\left(z\right)}{dz}
 \cdot
 \frac{ 1 } { \left(1+z\right)^2 }
 \cdot
 \sum_{\beta = e,\mu\,\tau}
 P_{\beta\alpha}\
 J_{\nu_\beta} \left[ E_0\left(1+z\right) \right] \;,  \label{equ:DiffuseFlux}
\end{equation}
where $E_0$ is the received neutrino energy, \ie, at $z=0$.  Here, $\rho_\text{src} \left( z \right)$ is the source number density (in units of cm$^{-3}$), the comoving distance to the source (in units of cm) is 
\begin{equation}
 r \left( z \right) = \int_0^z \frac { c } { H\left(w\right) } dw \;,
\end{equation}
and the differential comoving volume (in units of cm$^3$) is
\begin{equation}
 \frac { dV } { dz }
 =
 4 \pi
 \frac { c } { H\left(z\right) } r^2 \left( z \right ) \;,
\end{equation}
with $H\left(z\right) = H_0 \sqrt{ \Omega_m \left(1+z\right)^3 + \Omega_\Lambda } \equiv H_0 h \left( z \right)$ the Hubble parameter, $H_0$ the Hubble constant, and $\Omega_m$, $\Omega_k$, $\Omega_\Lambda$ the adimensional energy densities of matter, curvature, and cosmological constant.  If neutrinos decay, the probability $P_{\beta\alpha}$ of the flavor transition $\nu_\beta \to \nu_\alpha$ depends on neutrino energy, neutrino lifetime $\tau_i/m_i$, and source redshift, as shown in \App\ \ref{appendix:FlavorTransition}.  After simplification, \equ{DiffuseFlux} becomes
\begin{equation}
 E_0^2 \Phi_{\alpha}\hspace{-0.1cm}\left(E_0\right)
 =
 \frac { L_H } { 4 \pi }
 \int_0^\infty dz \, 
 \frac{ \rho_\text{src} \left(z\right) } { h \left(z\right) \left( 1+z \right)^2 } 
 \cdot
 \sum_{\beta = e,\mu\,\tau}
 P_{\beta\alpha}\
 J_{\nu_\beta} \left[ E_0\left(1+z\right) \right] \;,  \label{equ:DiffuseFlux2}
\end{equation}
where $L_H \equiv c / H_0 \approx 3.89$ Gpc is the Hubble length.

In our calculations, we assume that all flavors of neutrinos and anti-neutrinos have the same spectral shape.  Thus, we can write the luminosity of $\nu_\beta$ as
$J_{\nu_\beta}\hspace{-0.1cm}\left[ E_0\left(1+z\right) \right] = f_{\beta,\text{S}} J_{\nu_\text{all}}\hspace{-0.1cm} \left[ E_0\left(1+z\right) \right]$, with $f_{\beta,\text{S}}$ the flavor ratio of $\nu_\beta$ at the source and $J_{\nu_\text{all}}$ the all-flavor neutrino plus anti-neutrino luminosity. (We adopt here the same normalization condition as in \Sec\ \ref{section:DecayShowerRate}, $\sum_\alpha f_{\alpha,\text{S}} + f_{\bar{\alpha},\text{S}} = 1$).  Hence,
\begin{equation}
 \sum_{\beta = e,\mu\,\tau}
 P_{\beta\alpha}\
 J_{\nu_\beta} \left[ E_0\left(1+z\right) \right] 
 =
 \left(
 \sum_{\beta = e,\mu\,\tau}
 P_{\beta\alpha}\
 f_{\beta,\text{S}}
 \right)
 J_{\nu_\text{all}} \left[ E_0\left(1+z\right) \right]
 =
 f_{\alpha,\oplus}
 J_{\nu_\text{all}} \left[ E_0\left(1+z\right) \right] \;,
\end{equation}
and \equ{DiffuseFlux2} becomes
\begin{equation}
 E_0^2 \Phi_{\alpha}\hspace{-0.1cm}\left(E_0\right)
 =
 \frac { L_H } { 4 \pi }
 \int_0^\infty dz \, 
 \frac{ \rho_\text{src} \left(z\right) } { h \left(z\right) \left( 1+z \right)^2 } 
 \cdot
 f_{\alpha,\oplus}
 J_{\nu_\text{all}} \left[ E_0\left(1+z\right) \right] 
 \;,  \label{equ:DiffuseFlux3}
\end{equation}
with $f_{\alpha,\oplus}$ the flavor ratios at Earth.  The normalization of $E_0^2 \Phi_{\alpha}\hspace{-0.1cm}\left(E_0\right)$ is fixed by fitting it to reported IceCube fluxes.

Figure \ref{fig:differential-flux-vs-redshift} shows the differential diffuse flux, \equ{DiffuseFlux3}, as a function of redshift, for two choices of spectral index (and assuming no decay).  For concreteness, in this plot and all of our results, we have assumed that the luminosity density $\rho_\text{src} \cdot J_{\nu}$ follows the star formation rate\ \cite{Hopkins:2006bw,Yuksel:2008cu}.  Clearly, the sources that contribute the most to the diffuse flux lie between $z \approx$ 0.5--1, or $r \approx$ 2--3 Gpc.


\section{Shower rate}\label{section:ShowerRate}

\setcounter{figure}{0}
\renewcommand{\thefigure}{C\arabic{figure}}
\begin{figure}[t!]
 \centering
 \includegraphics[width=0.5\textwidth]{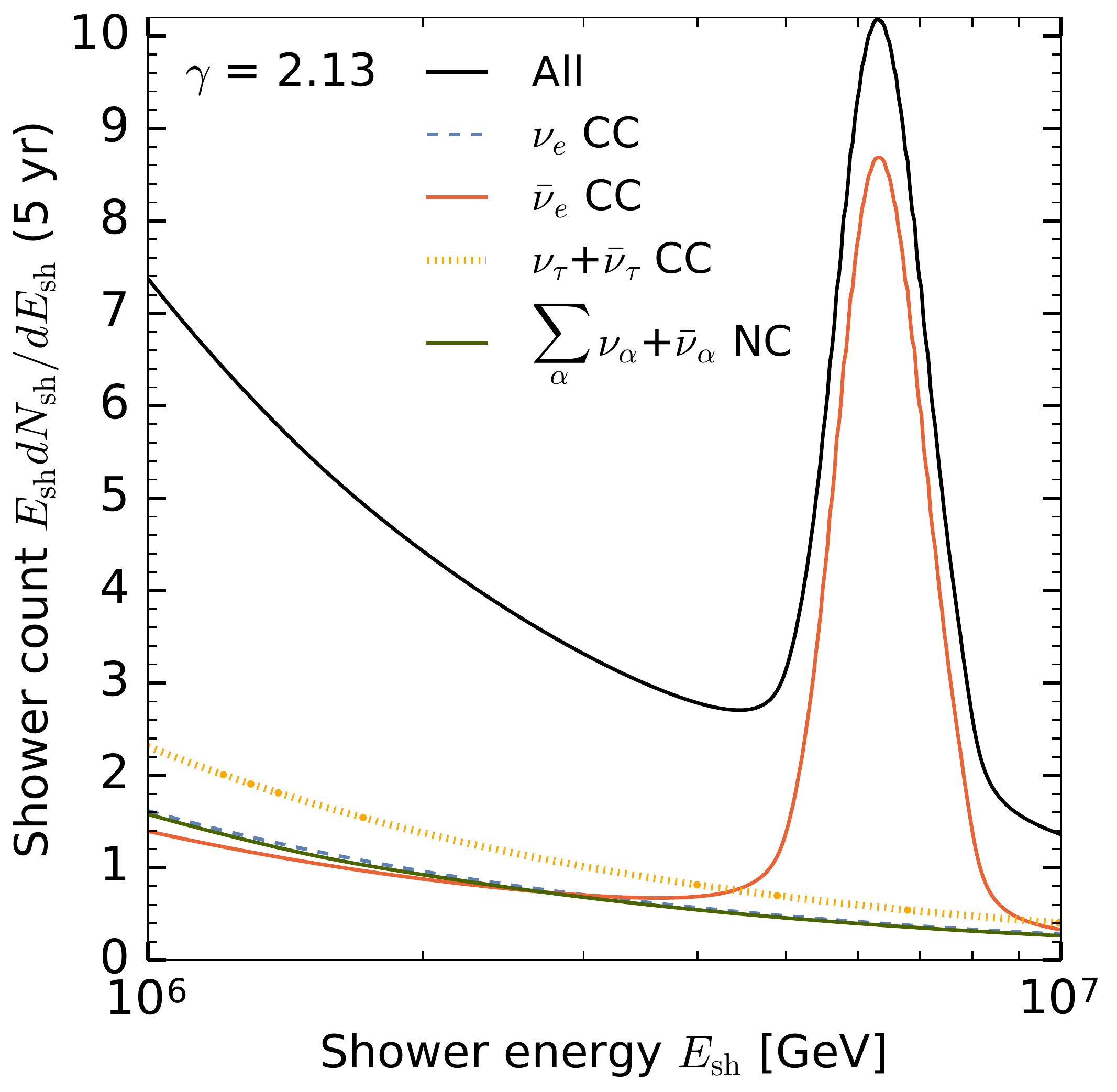}
 \caption{Components of the shower spectrum at IceCube, assuming a flux $\propto E^{-2.13}$\ \cite{Aartsen:2016xlq} and five years of exposure.  The detector energy resolution is set to $\delta E_\text{sh} / E_\text{sh} = 0.1$\ \cite{Aartsen:2013vja}. See text for details.}
 \label{fig:shower-rates-components}
\end{figure}

High-energy neutrinos undergo deep-inelastic scattering off nucleons in the Antarctic ice.  Charged-current (CC) interactions create charged leptons: $\nu_l + N \to l + X$ ($l = e, \mu, \tau$), where $X$ are final-state hadrons.  Neutral-current (NC) interactions create neutrinos: $\nu_l + N \to \nu_l + X$.  Outgoing hadrons carry a fraction $y$ of the neutrino energy, while leptons carry $\left( 1- y \right)$, where $y$ is the inelasticity.  IceCube PMTs collect the Cherenkov light produced by the final-state hadrons and charged leptons. 

The muon from a $\nu_\mu$ CC interaction leaves a track of light a few kilometers long that is identifiable.  Tracks also come from the decay of taus, produced in $\nu_\tau$ CC interactions, into muons, which occurs 17\% of the time.

All other outgoing particles, except neutrinos, create particle showers localized around the interaction vertex.  Final-state hadrons create a shower with high neutron and pion content --- a hadronic shower.  In a NC interaction, this is the only shower.  In a $\nu_e$ CC interaction, the electron creates an additional shower that contains mostly electrons, positrons, and gamma rays, with few hadrons --- an electromagnetic shower.  In a $\nu_\tau$ CC interaction, the tau creates a hadronic shower 66\% of the time and an electromagnetic shower 17\% of the time. The lepton- and hadron-initiated showers in CC interactions are not resolved individually; their superposition is recorded as a single shower.

Shower detection in IceCube is calorimetric: the energy of the particles that initiated the shower can be reconstructed closely from the collected light. The relation between neutrino energy $E_\nu \equiv E_0$ and shower energy $E_\text{sh}$ depends on flavor and interaction type. In a $\nu_e$ CC interaction, all of the neutrino energy is deposited in showers. In a $\nu_\tau$ CC interaction, about $30\%$ of the tau energy is lost to neutrinos at decay.  Since $17\%$ of tau decays are into muons and neutrinos, only $83\%$ of $\nu_\tau$ CC interactions create showers.  NC interactions only deposit, on average, $\langle y \rangle E_\nu$ as hadronic showers.  Around 1 PeV, it is $\langle y \rangle \approx 0.25$ for neutrinos and anti-neutrinos, and for CC and NC\ \cite{Gandhi:1995tf}. So, the NC contribution to the total shower rate is sub-dominant. In summary,
\begin{equation}\label{equ:ShowerNeutrinoEnergy}
 E_\text{sh} \simeq
 \left\{\begin{array}{ll}
  E_\nu                                                                                              & \text{for } \nu_e \text{ CC} \\
  \left[ \langle y \rangle + 0.7 \left( 1 - \langle y \rangle \right) \right] E_\nu \simeq 0.8 E_\nu & \text{for } \nu_\tau \text{ CC} \\
  \langle y \rangle E_\nu \simeq 0.25 E_\nu                                                          & \text{for } \nu_x \text{ NC}
 \end{array}\right. \;.
\end{equation}
(See also \Ref\ \cite{Blum:2014ewa}, where different decay modes of the tau are treated separately.)

To calculate the energy spectrum of showers in IceCube, we follow the ``theorist's approach'' from \Ref\ \cite{Laha:2013lka} (see also \Ref\ \cite{Blum:2014ewa}):
\begin{equation}\label{equ:dNdEcasc}
 \frac{dN}{dE_\text{sh}}
 = \frac{dN_e^\text{CC}}{dE_\text{sh}} + 0.83 \cdot \frac{dN_\tau^\text{CC}}{dE_\text{sh}} + \sum_{\alpha=e,\mu,\tau} \frac{dN_\alpha^\text{NC}}{dE_\text{sh}}
\end{equation}
with
\begin{eqnarray}\label{equ:dNdEcascPerFlavor}
 \frac{dN_{\alpha}^j}{dE_\text{sh}} \left( E_\text{sh} \right)
 \simeq
 2 \pi \rho_\text{ice} N_\text{A} V T
 \int_{-1}^{+1} d\left(\cos \theta_z\right)
 \left( \frac{d\Phi_\alpha}{dE_\nu}\left(E_\nu\right) \sigma_{\nu N}^j\left(E_\nu\right) e^{-\tau_\alpha\left(E_\nu,\cos \theta_z\right)} \,+ \right. \nonumber \\
 \left. \frac{d\Phi_{\bar{\alpha}}}{dE_\nu}\left(E_\nu\right) \left( \sigma_{\bar{\nu} N}^j\left(E_\nu\right) + \delta_{j,\text{CC}} \delta_{\alpha e} \sigma_{\bar{\nu}_e e}^\text{CC} \left(E_\nu\right)  \right) e^{-\tau_{\bar{\alpha}}\left(E_\nu,\cos \theta_z\right)} \right) \;, 
\end{eqnarray}
where $j = $ CC (charged current) or NC (neutral current).  The diffuse flux of $\nu_\alpha$ is calculated in \equ{DiffuseFlux}.  On the right-hand side of \equ{dNdEcascPerFlavor}, the neutrino energy in the integrand is calculated from the provided value of shower energy, via \equ{ShowerNeutrinoEnergy}.  The number of target nucleons is $\rho_\text{ice} N_\text{A} V$, with $\rho_\text{ice} \approx 0.92$ g cm$^{-3}$ the density of ice, $N_\text{A}$ the Avogadro number, and $V \approx 1$ km$^{3}$ the volume of IceCube.  The volume is constant at high energies (see \Fig\ 7 in \Ref\ \cite{Aartsen:2013jdh}); we assume that optimized HESE cuts would not reduce this volume appreciably.  In the main text, we set the exposure time to $T = 5$ yr.  

The shower rate calculated in \equ{dNdEcasc} includes the $\nu_\tau$ CC contribution.  However, in the energy interval 5--8 PeV considered in the main text, these interactions should be separately identifiable as double bangs\ \cite{Learned:1994wg}, lollipops\ \cite{Beacom:2003nh}, or tau-to-muon decays \cite{DeYoung:2006fg} at high energies, and as double pulses at low energies\ \cite{Aartsen:2015dlt}; thus, they would not contribute appreciably to the shower rate.  Accordingly, Figs.\ \ref{fig:cascade-rates} and \ref{fig:cascades-vs-lifetime} do not contain the $\nu_\tau$ CC contribution, \ie, the term $dN_\tau^\text{CC}/dE_\text{sh}$ is suppressed in them.

A neutrino with incoming zenith angle $\theta_z$ traverses a distance
\begin{equation}
 l = \sqrt{ \left( R_\oplus^2 - 2 R_\oplus d \right) \cos^2 \theta_z + 2 R_\oplus d } - \left( R_\oplus - d \right) \cos \theta_z
\end{equation}
inside the Earth, which has radius $R_\oplus$, before reaching a detector that is buried a distance $d$ below the surface.  For IceCube, $d \approx 1.5$ km.  For each incoming direction, we calculate the average Earth density $\langle \rho_\oplus \rangle = (1/l) \int_0^l \rho_\oplus\left(x\right) dx$, with the local density $\rho_\oplus$ given by the Preliminary Earth Reference Model\ \cite{Dziewonski:1981xy,Gandhi:1995tf}.

We use the neutrino-nucleon cross sections $\sigma_{\nu N}^\text{CC}$ and $\sigma_{\nu N}^\text{NC}$ from \Ref\ \cite{Connolly:2011vc}; at these energies, the cross sections for neutrinos and anti-neutrinos are similar.  We use the Glashow resonance cross section $\sigma_{\bar{\nu}_e e}^\text{CC}$, with decay of the $W$ boson into hadrons, as calculated in \Ref\ \cite{Gandhi:1995tf}.  The mean free path of the neutrino inside the Earth is
\begin{equation}
 \lambda_{\alpha} =
 \frac{ m_N }{ \langle \rho_\oplus \rangle} \left( \frac {1} {{\sigma_{\nu N}^\text{CC}} +\sigma_{\nu N}^\text{NC}} \right) \;,
\end{equation}
for $\nu_e$, $\nu_\mu$, $\nu_\tau$; $\lambda_{\bar{\alpha}} \equiv \lambda_\alpha \left( \nu \to \bar{\nu} \right)$, for $\bar{\nu}_\mu$, $\bar{\nu}_\tau$; and
\begin{equation}
 \lambda_{\bar{e}} =
 \frac{ m_N }{ \langle \rho_\oplus \rangle} \left( \frac {1} {{\sigma_{\bar{\nu} N}^\text{CC}} +\sigma_{\bar{\nu} N}^\text{NC} + \langle Y_e \rangle \sigma_{\bar{\nu}_e e}^\text{CC}} \right) \;,
\end{equation}
for $\bar{\nu}_e$, where $\langle Y_e \rangle \approx 0.5$ is the average number of electrons per nucleon in the Earth.  The optical depth, which accounts for attenuation of the flux inside the Earth, is then calculated as $\tau_{\alpha} = l/\lambda_{\alpha}$.

The detector energy resolution has been taken into account by folding \equ{dNdEcasc} with a Gaussian of width $\delta E_\text{sh} / E_\text{sh} = 0.10$, consistent with the value reported by IceCube\ \cite{Aartsen:2013vja}.

Figure\ \ref{fig:shower-rates-components} shows, for illustration purposes, the shower spectrum divided into its contributing components, by flavor and interaction type.  The contribution from the Glashow resonance is clearly dominant in the range 5--8 PeV.

\end{document}